\documentclass[sigconf, screen]{acmart}

\setcopyright{none}
\renewcommand\footnotetextcopyrightpermission[1]{}

\usepackage{subcaption}
\usepackage{multirow}
\usepackage{pifont}
\usepackage{siunitx}
\usepackage{booktabs}
\usepackage{array}
\usepackage{tabularx}
\usepackage{makecell}

\newcommand{\ToolName}{Pudding}

\newcounter{contribhead}
\newcommand{\contribhead}[1]{\refstepcounter{contribhead}\textit{\alph{contribhead})~#1.}\ }

\newcommand{\Cthree}{C\textsuperscript{3}}

\keywords{Processing-Using-DRAM, compilers, MLIR, equality saturation, integer linear programming}

\begin{document}

\title{\ToolName{}: A Retargetable Compiler for Processing-Using-DRAM Systems}

\author{Tianrui Zheng}
\authornote{Both authors contributed equally.}
\authornote{Work partially done at TU Dresden.}
\email{zhengtr1998@gmail.com}
\affiliation{\institution{TU~Darmstadt, Huawei~Hilbert~Research~Center}\country{Germany}}

\author{Jo\~{a}o Paulo Cardoso de Lima}
\authornotemark[1]
\authornotemark[2]
\email{joao.lima@fau.de}
\affiliation{\institution{Friedrich-Alexander-Universit\"{a}t Erlangen-N\"{u}rnberg}\country{Germany}}

\author{Alexander Kusnezoff}
\email{alexander.kusnezoff@mailbox.tu-dresden.de}
\affiliation{\institution{TU Dresden}\country{Germany}}

\author{Asif Ali Khan}
\authornotemark[2]
\email{asif.khan@uni-paderborn.de}
\affiliation{\institution{Paderborn University}\country{Germany}}

\author{Jeronimo Castrillon}
\email{jeronimo.castrillon@tu-dresden.de}
\affiliation{\institution{TU Dresden, ScaDS.AI}\country{Germany}}

\begin{abstract}
Processing-Using-DRAM (PUD) architectures exploit the physics of DRAM arrays to perform Boolean operations directly in memory, eliminating the data movement bottleneck in memory-intensive workloads.
Several PUD architectures have been recently proposed which differ in their primitive operation sets, row allocation models, and execution metrics.
These architectural differences are today hard-coded in emerging PUD-aware compilers, making cross-architecture reuse impossible.
We present a retargetable, end-to-end compilation framework for PUD architectures. The central idea is to make the architecture description a first-class object in the compilation pipeline: retargeting to a new PUD architecture only requires supplying its description. Built upon the MLIR framework, our compiler operates at two coupled levels. At the tensor level, it lowers high-level programs through bit-slicing, mixed-precision assignment, and integer-linear-programming (ILP) based placement to reduce the number of PUD commands and inter-bank data movement. At the logic level, it employs equality saturation to explore Boolean network implementations, with rewrite rules and cost functions derived directly from the architecture description. Evaluated on representative workloads across multiple PUD targets with the same compiler passes, our framework matches (and in several cases surpasses) architecture-specific hand-optimized compilers, while delivering retargetability and extensibility.

\end{abstract}

\maketitle
\pagestyle{plain}

\section{Introduction}
\label{sec:intro}

Recent experimental work confirms that commercial off-the-shelf (COTS) DDR4 chips already support a functionally-complete set of Boolean primitives (i.e., \texttt{NOT} and multi-input \texttt{AND}, \texttt{OR}, and \texttt{MAJ}) without hardware modifications~\cite{gao2019computedram,gao2022fracdram,yuksel2023pulsar,yuksel2024functionally,he2026page,ji2026c3}. Industry has also begun to publish PUD work, e.g., SK hynix~\cite{he2026page,ji2026c3}. Other designs propose small modifications to the DRAM array to support a richer set of primitives and enable more efficient computation in memory~\cite{seshadri2017ambit,li2017drisa,angizi2019redram,hajinazar2021simdram,zhou2022flexidram,shin2024processing,oliveira2024mimdram,de2025proteus,de2026count2multiply,oliveira2026pipedram}.

Each design, however, comes with its own mapping constraints, which impose substantially different requirements on code generation. For example, a SIMDRAM-class target~\cite{hajinazar2021simdram} exposes \texttt{MAJ3}, \texttt{NOT}, and \texttt{RowCopy} as its native primitives and separates compute rows from data rows within the same subarray. Such a target requires a logic optimizer and a row allocator built around these assumptions. In contrast, a COTS DDR4 target~\cite{yuksel2023pulsar,yuksel2024functionally} exposes a wider primitive set with primitive-specific row-usage rules. The same program therefore requires fundamentally different implementations on the two targets, and today each target's mapping is derived separately~\cite{guo2025pimsynth,peng2023chopper}. As a result, porting an application to a new target means repeating this mapping, and comparing targets on the same application requires repeating it for every target.

A PUD design that is still evolving cannot be evaluated on full applications until a compiler targets it, as also observed for digital PIM~\cite{noor2026pimautodse}. Compiler techniques have often been developed ahead of the hardware they target, as with trace scheduling for VLIW machines~\cite{fisher1981trace}, compilers for quantum programs~\cite{javadiabhari2014scaffcc}, and data placement for racetrack memories~\cite{khan2019shiftsreduce}.
Mapping a high-level program onto a PUD architecture requires solving optimization problems at two levels. At the \emph{tensor level}, the compiler must decide how operations are ordered and fused, how tensors are tiled and partitioned across the memory banks and subarrays, and what bitwidth each operator requires. Bitwidth optimizations are specifically crucial because the number of PUD commands issued for an operation scales directly with operand bitwidth.
At the \emph{logic level}, the compiler must decompose each arithmetic operation into a sequence of target-specific atomic operation (primitives), map operand bits onto memory rows, and generate an optimal command sequence that realizes the computation.
A \emph{key observation} is that these two levels are dependent in ways that do not arise in conventional compilation. The bitwidth decisions made at the tensor level, for instance, determine the row allocation of the Boolean networks that the logic optimizer must handle, and the layout constraints at the logic level feed back into how tensors can be partitioned at the tensor level. A change in the target architecture (e.g., primitive set, row budget, and row decoding scheme) propagates through both levels simultaneously, which is why a retargetable framework must treat them jointly rather than as independent phases.

Existing frameworks hard-code architectural constraints directly into their backend, making cross-target reuse structurally impossible~(\autoref{tab:related}).
SIMDRAM and its successors use \texttt{MAJ}/\texttt{NOT} as the canonical basis and build compilation pipelines atop it~\cite{hajinazar2021simdram,oliveira2024mimdram,de2025proteus,kubo2025mvdram,de2024count2multiply}. CINM~\cite{khan2024cinm} introduces a common intermediate representation (IR) and generic optimization infrastructure. However, the technology-specific backend, i.e., the logic optimization, layout, command generation, are explicitly unspecified and unshared across targets.
The backends are meant to be provided by the HW vendors. Other tools (e.g., Primo, Chopper, OptiPIM, PIMSynth and Meta-PIM~\cite{PRIMO,peng2023chopper,liu2025optipim,guo2025pimsynth,zhang2026meta}) target isolated subproblems with target-specific heuristics and no shared abstraction integrating logic and layout. The result is a fragmented ecosystem where every new PUD architecture demands a new compiler, and common optimizations must be reimplemented from scratch. This fragmentation stems from a deeper misconception, i.e., existing tools treat PUD as an emulator of a fixed operation set rather than as a programmable logic substrate, a critical distinction when arithmetic is non-standard and optimal Boolean implementations must be synthesized against target-specific primitive costs.
Compounding this, there is neither a \emph{common abstraction} that can capture cross-architecture features such as bit-slicing, destructive operation semantics, and layout orientation, nor a retargetable \emph{mapping framework} that jointly optimizes across tensor and logic levels under architecture-specific constraints.

\newcommand{\cmark}{\ding{51}}
\newcommand{\xmark}{\ding{55}}
\newcommand{\pmark}{$\circ$}
\begin{table}[t]
\scriptsize
\centering
\caption{Hardware frameworks vs.\ compilers for in-memory computing. \cmark: supported; \pmark: partial; \xmark: not supported.
MT: multiple targets; Synth.: synthesizes arbitrary operations into target primitives; Row: row allocation; Destr.: models destructive operations; Tensor: tensor-level mapping/scheduling; Bitw.: bitwidth analysis; Input: programming entry point.}
\label{tab:related}
\setlength{\tabcolsep}{2.6pt}
\begin{tabular*}{\columnwidth}{@{\extracolsep{\fill}}lccccccl@{}}
\toprule
\textbf{Work} & \textbf{MT} & \textbf{Synth.} & \textbf{Row} & \textbf{Destr.} & \textbf{Tensor} & \textbf{Bitw.} & \textbf{Input} \\
\midrule
\multicolumn{8}{@{}l}{\textit{PUD HW frameworks (ISA + $\mu$Program library)}} \\
SIMDRAM~\cite{hajinazar2021simdram}  & \xmark & \pmark & \cmark & \cmark & \xmark & \xmark & ISA \\ %
Proteus~\cite{de2025proteus}         & \xmark & \xmark & \cmark & \cmark & \xmark & \cmark & ISA \\ %
\midrule
\multicolumn{8}{@{}l}{\textit{Logic-level PUD compilers}} \\
CHOPPER~\cite{peng2023chopper}       & \cmark & \cmark & \cmark & \cmark & \pmark & \xmark & Usuba \\ %
PIMsynth~\cite{guo2025pimsynth}      & \pmark & \cmark & \cmark & \cmark & \xmark & \xmark & Verilog \\ %
\midrule
\multicolumn{8}{@{}l}{\textit{Tensor-level PIM compilers}} \\
OptiPIM~\cite{liu2025optipim}        & \cmark & \xmark & \xmark & \xmark & \cmark & \xmark & PyTorch \\ %
Meta-PIM~\cite{zhang2026meta}        & \pmark & \xmark & \xmark & \xmark & \cmark & \xmark & C \\ %
CINM~\cite{khan2024cinm}             & \cmark & \xmark & \xmark & \xmark & \cmark & \xmark & MLIR \\ %
OCC~\cite{occ}                       & \xmark & \xmark & \xmark & \xmark & \cmark & \xmark & TC \\ %
CIM-MLC~\cite{qu2024cim}             & \cmark & \xmark & \xmark & \xmark & \cmark & \xmark & ONNX \\ %
\midrule
\textbf{\ToolName{}}                 & \cmark & \cmark & \cmark & \cmark & \cmark & \cmark & PyTorch \\
\bottomrule
\end{tabular*}%
\end{table}

We present \ToolName{}, the first retargetable, end-to-end compiler for PUD architectures.
\ToolName{}'s central idea is to make the architecture description a first-class object that propagates through every compilation stage, such as  parameterizing rewrite rules, primitive costs, data layout constraints, and mapping and scheduling decisions, so that retargeting requires updating a description file rather than rewriting compiler passes. Our concrete contributions are as follows:

\contribhead{A parameterized primitive model} We define an architecture description for PUD systems that captures a target's supported compute primitives, their command costs, compute row budgets per subarray, and bank and subarray structural constraints.
This description serves as the shared interface across compilation stages.
We instantiate it for both modified and unmodified COTS DRAM targets, demonstrating that qualitatively different architectures are expressible within the same framework.

\contribhead{An architecture-aware logic optimizer and code generator} We develop a logic optimizer that uses equality saturation over e-graphs~\cite{tate2009equality} to rewrite and select Boolean network implementations, with rewrite rules derived automatically from the target's primitive model. Critically, the optimizer explores subnetwork equivalences
rather than committing to a fixed representation: the e-graph maintains all equivalent implementations, from which the one minimizing command count on the target is extracted.

\contribhead{A mapping and scheduling optimizer} We present a joint mapping and scheduling approach for executing tensor programs on multi-bank DRAM architectures.
Our ILP formulation captures the trade-off between parallelism and inter-bank communication. Compared to the state of the art~\cite{liu2025optipim}, our approach not only incorporates bank-level resource constraints, but also models operation-specific latency, %
capturing differences across arithmetic and logic operations. The resulting mapper jointly optimizes operation placement and data layout to minimize end-to-end execution time while ensuring correct instruction dispatch and execution.

\contribhead{An end-to-end MLIR-based framework} We develop a compiler that ingests high-level arithmetic programs from ML frameworks, progressively lowers through an extensible MLIR dialect flow to PUD primitives, and emits optimized command sequences for a specified target. The integration is the central contribution: for the first time, architecture-awareness flows coherently from primitive modeling through logic optimization through layout without per-target manual intervention.

Across three PUD architectures, \ToolName{} delivers up to 1.92\(\times\) cost efficiency in logic-level optimizations and up to 3.25\(\times\) speedup over decoupled mapping in tensor-level optimizations, while achieving 1.06--1.38\(\times\) mapping speedup over OptiPIM~\cite{liu2025optipim} with valid layouts at every kernel size and bank count.
On vector search over SIFT1M~\cite{jegou2011SIFT1M}, compiling end-to-end from PyTorch above the kernel level raises the speedup over the CPU baseline from $2.9\times$ to $5.8\times$.

\section{Background}
\label{sec:related}

\subsection{DRAM Organization and Operation}
\label{subsec:dram-CIM}
\begin{figure}[tbp]
\centering
\includegraphics[width=0.95\columnwidth]{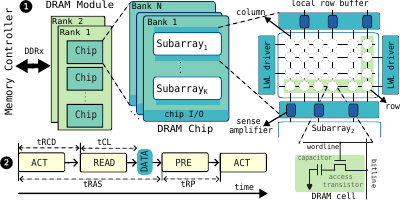}
\caption{Overview of DRAM organization.}
\label{fig:dram}
\end{figure}

\autoref{fig:dram} illustrates the hierarchical organization of a modern DRAM system. At the system level (\ding{182}), a CPU issues memory requests through one or more memory controllers, each managing an independent channel connected to one or more DRAM modules. Each module contains rank groups of DRAM chips operating in lockstep to collectively supply a full data bus width (typically 64 bits). Within each chip, memory is partitioned into banks sharing an internal data bus to the chip's I/O circuitry, and each bank is further divided into subarrays, where storage occurs in a two-dimensional array of DRAM cells, each a capacitor and access transistor. Peripheral circuitry within each subarray, i.e., local wordline (LWL) drivers and sense amplifiers (SAs), orchestrates data access and restoration.

\autoref{fig:dram} (\ding{183}) illustrates the command sequence for a DRAM read. First, the controller issues an activate command (\texttt{ACT}) with the target row address; the LWL decoder drives the corresponding wordline high, connecting the row's cells to their bitlines, and the sense amplifiers detect and amplify the small charge differential, latching the row into the row buffer. Second, the controller issues a \texttt{READ} command with a column address; the column decoder selects the appropriate sense amplifiers, and the data propagates through the internal bus to the chip's I/O interface and onto the channel. Finally, the controller issues a precharge command (\texttt{PRE}), restoring the bitlines to $V_{DD}/2$ and returning the bank to the idle state, ready for the next activation.

\begin{figure*}[tbp]
\includegraphics[width=0.90\textwidth]{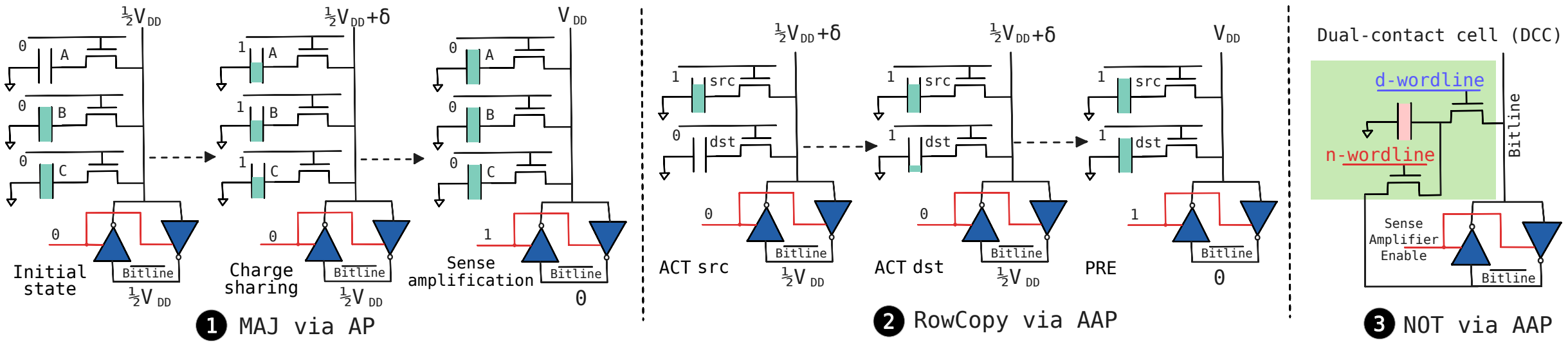}
\centering
\caption{Overview of Processing-Using-DRAM operations.}
\label{fig:pud}
\end{figure*}

\subsection{Processing-Using-DRAM (PUD)}
\label{sec:background:PUD}

PUD exploits standard DRAM commands to compute directly in the array. RowClone~\cite{seshadri2013rowclone} (\autoref{fig:pud}, \ding{183}) uses an \emph{activate-activate-precharge} (\texttt{AAP}) sequence: \texttt{ACT} on the source latches it into sense amplifiers, \texttt{ACT} on the destination triggers charge-sharing that overwrites it with the source data, then \texttt{PRE} resets the bitlines. Building on this, \emph{multi-row activation} (\texttt{MRA}) activates several rows in a subarray at once followed by precharge (an \emph{activate-precharge}, \texttt{AP}, sequence) to perform bulk bitwise logic~\cite{seshadri2017ambit,hajinazar2021simdram,gao2019computedram}: these \texttt{AAP}/\texttt{AP} sequences are called \emph{primitives}, the basic building blocks of PUD. An example is \texttt{MAJ3} (\autoref{fig:pud}, \ding{182}), where charge sharing across three rows yields a majority-encoded voltage deviation that the sense amplifier resolves into a digital result, destructively overwriting all three rows. \texttt{MRA} can be achieved via custom row decoders asserting multiple wordlines at once~\cite{seshadri2017ambit}, or by violating DRAM timing constraints with back-to-back \texttt{ACT} commands so multiple wordlines stay active simultaneously~\cite{gao2019computedram,shin2024processing,yuksel2023pulsar}.

For functional completeness, \texttt{NOT} is implemented either via dual-contact cells (DCCs) (\autoref{fig:pud}, \ding{184}), which are cells wired to the bitline or its complement through dedicated wordlines, populated via \texttt{RowCopy}~\cite{seshadri2017ambit}, or, in PRADA~\cite{shin2024processing}, by exploiting internal sense-amplifier behavior (OCSA-type, unlike standard BLSAs) to invert data with no extra circuitry. Similarly, different row allocation strategies for computing within the array exist: reserving dedicated rows at design time for simpler, more reliable sensing at the cost of storage capacity and required manufacturer support~\cite{seshadri2017ambit}, versus letting arbitrary rows compute via timing violations~\cite{gao2019computedram,shin2024processing}, thus recovering capacity at the cost of circuit complexity or generality. Finally, unmodified COTS DRAM can still perform Boolean logic (\texttt{AND}/\texttt{OR}) by activating rows in neighboring subarrays sharing sense amplifiers: an \emph{activate-precharge-activate} (\texttt{APA}) sequence uses reference rows pre-set to fractional voltages~\cite{gao2022fracdram} to bias the shared amplifiers while adjacent operand rows drive the actual computation~\cite{gao2019computedram,yuksel2024functionally}.

\subsection{The MLIR Compiler Infrastructure}
\label{subsec:MLIR}
MLIR is a compiler infrastructure for representing and transforming programs across multiple abstraction levels and heterogeneous hardware targets~\cite{lattner2021mlir}.
Instead of a fixed intermediate representation (IR), MLIR provides a small set of builtins and leaves the rest extensible through \emph{dialects}, i.e., logical groupings of custom types, operations, and attributes. Operations are the core building blocks, consuming and producing typed values known at compile time and carrying compile-time metadata via attributes. Dialects are typically organized by domain (\texttt{linalg}, \texttt{tosa}), representation (\texttt{affine}, \texttt{scf}, \texttt{memref}), or target (\texttt{gpu}, \texttt{x86vector}).
Abstractions are progressively \emph{lowered} from high-level, domain-specific dialects toward platform-specific ones through dialect conversion passes, and can also be \emph{raised} to recover higher-level structure~\cite{brauckmann2023mlirsynth}. Each dialect preserves the preconditions for its transformations directly in the IR, reducing analysis cost and complexity. \emph{Pattern rewriting}, i.e., replacing operations that match a given IR pattern with semantically equivalent ones at a different abstraction level, is the primary vehicle for both lowering and optimization exposed through a composable, greedy rewrite engine.

\subsection{Equivalence Graphs for Boolean Rewriting}
\label{sec:background:egraph}
Equivalence graphs (e-graphs) are a data structure for compactly representing large sets of equivalent terms. An e-graph contains a set of \emph{equivalence classes} (e-classes), each holding a set of \emph{e-nodes}; every e-node consists of a function symbol and an ordered list of child e-classes.
The key operation on e-graphs is \emph{equality saturation}~\cite{tate2009equality, willsey2021egg}: starting from an initial term, a set of rewriting rules $\ell \rightarrow r$ is applied repeatedly. Whenever a term matching $\ell$ is found, $r$ is added to the same e-class as $\ell$ rather than replacing it. This continues until no new rewrites apply (saturation) or a stopping condition, such as a timeout, is reached.

This non-destructive rewriting is the central advantage of e-graphs over traditional term rewriting~\cite{chen2024syn,hofmann2025eqmap}, which must commit to each rewrite greedily. For example, applying associativity to $\texttt{MAJ}(a, b, \texttt{MAJ}(a, c, c))$ yields $\texttt{MAJ}(a, c, \texttt{MAJ}(a, b, c))$, after which the idempotence pattern $\texttt{MAJ}(x, x, y) = x$ is no longer visible at the top level, obscuring the optimal result $\texttt{MAJ}(a,b,c)$.
In contrast, an e-graph retains all intermediate and derived terms simultaneously, so the optimal form remains reachable regardless of rule-application order.

\section{Motivation}
\label{sec:motiv}
Retargeting a PUD compiler is not simply a matter of swapping ISAs. The axes of variation in PUD designs cut through multiple compiler layers simultaneously. The primitive set determines how Boolean networks must be structured; the row allocation model determines operand placement and row budget; destructive operation semantics constrain liveness; and data layout orientation governs how tensor dimensions map onto subarray rows. A change in any one property invalidates decisions made at every other level. We illustrate this with two targets that are both bulk-bitwise DRAM systems, yet require fundamentally different compilation strategies.

\begin{figure}[tbp]
\centering
\includegraphics[width=0.90\columnwidth]{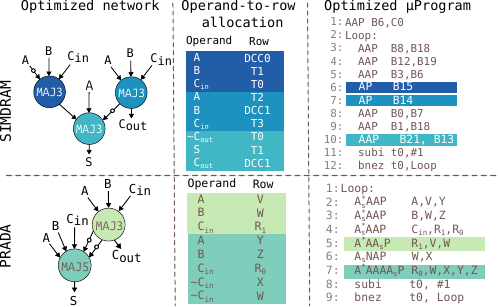}
\caption{1-bit full adder in SIMDRAM~\cite{hajinazar2021simdram} and PRADA~\cite{shin2024processing}.}
\label{fig:motiv}
\end{figure}

Let us consider the 1-bit full adder example in \autoref{fig:motiv}.
On a SIMDRAM-class target~\cite{hajinazar2021simdram}, the primitive set (\texttt{AAP}/\texttt{AP}) naturally realizes \texttt{RowCopy}, \texttt{NOT} and \texttt{MAJ3} with designated rows for \texttt{MRA} and DCCs. SIMDRAM's framework converts an \texttt{AND}/\texttt{OR}/\texttt{XOR}/ \texttt{NOT} representation into an optimized Majority-Inverter Graph (MIG), allocates compute rows to \texttt{MAJ} node operands, and emits a sequence of \texttt{AAP}/\texttt{AP}. The entire optimization pipeline (i.e., logic optimizer, row allocator, code generator) is inseparable from those assumptions.
On PRADA~\cite{shin2024processing}, the primitives (e.g., \texttt{AAAP} and \texttt{ANAP}) are command sequences consisting of one \texttt{ACT} (\texttt{A}) per activated row plus an optional NOT command (\texttt{N}), terminated by \texttt{PRE} (\texttt{P}).
By relaxing both row-role rigidity and sequential activation, PRADA enables operations with more than three operands (e.g., \texttt{MAJ5}), in-place operand consumption, and a fundamentally different allocation strategy.
The same full adder compiles to structurally different Boolean networks, different row allocations, and non-equivalent command sequences with no portability between them at any level.

These divergences do not stay at the logic level; \emph{architectural variations propagate upward}. \textbf{Consequence 1}: Tensor-level mapping cannot be optimized independently of the lower-level arithmetic costs. In bit-serial execution, command count scales linearly with operand bitwidth for addition and quadratically for multiplication, so reducing precision by even one bit saves commands across every operation in the Boolean network. Tensor-level decisions, such as mixed-precision assignment and operation scheduling, in turn, have an amplified effect, and must be informed by the target's bit-level costs to be meaningful. OptiPIM~\cite{liu2025optipim} demonstrates this by showing that tensor-level mapping quality improves substantially when the ILP is informed by lower-level command costs.
\textbf{Consequence 2}: The arithmetic strategy determines the data layout. If the target's primitive set makes bit-parallel carry-save addition cheaper than bit-serial ripple-carry~\cite{kubo2025mvdram}, the compiler must adopt a horizontal data layout and insert explicit transposition wherever vertical layout is assumed. This decision cannot be made at the tensor level alone: it depends on the arithmetic strategy, which depends on the target's primitive costs. Layout is not a preprocessing step, it is a consequence of logic-level compilation, and it differs across targets.

\noindent\textbf{What is needed.} A retargetable PUD compiler must abstract over what designs share, such as bit-sliced layout, destructive and row-level operation semantics, while parameterizing what varies: primitive sets, row models, bit-serial vs. bit-parallel execution, and costs. Target-independent optimizations should be implemented once; target-dependent ones should be driven by the architecture description. We present the first compiler that satisfies these requirements.

\section{\ToolName{} Flow Overview}
\label{sec:overview}

\begin{figure}[tbp]
    \centering
    \includegraphics[width=\linewidth, trim=1.2cm 0.2cm 0.7cm 0.2cm, clip]{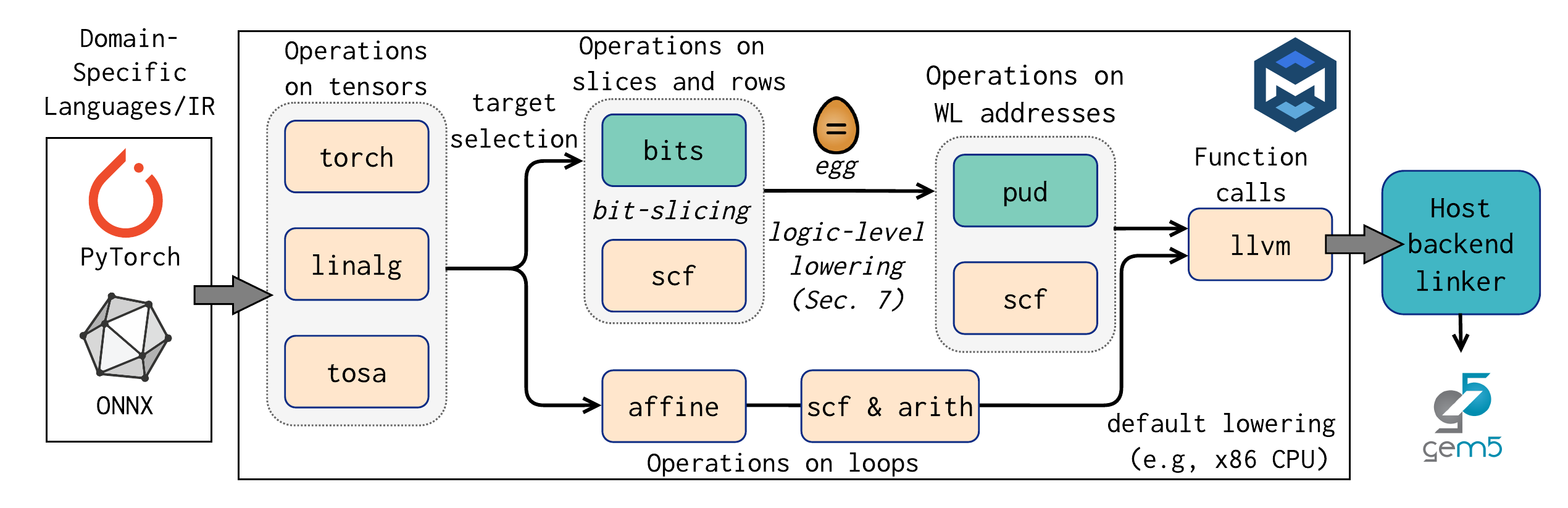}
    \caption{Overview of the MLIR-based \ToolName{} flow.}
    \label{fig:overview}
\end{figure}

\autoref{fig:overview} presents an overview of our multi-level compilation flow. \ToolName{} accepts programs from high-level frameworks such as PyTorch~\cite{pytorch} through torch-mlir~\cite{torchmlir2022} and partitions the resulting \texttt{linalg} program between PUD and the host CPU. An operator is offloaded to PUD if (a) it matches one of the computation patterns to be lowered to the \texttt{bits} dialect, i.e., a data-parallel computation with regular access; and if (b) its iteration space exceeds a configurable threshold that scales with the parallelism of the target memory configuration, by default the total number of SIMD lanes across its processing units. All other operators are lowered via MLIR's CPU pipeline.

For the offloaded operators, we first transform computations on \texttt{tensor} into computations on \texttt{slice}, a data type defined in our \texttt{bits} dialect (\autoref{sec:bits}). A \texttt{slice} is a two-dimensional bit-plane that maps one-to-one to a memory array.
Each \texttt{slice} is derived from a one-dimensional \texttt{tensor}, while higher-dimensional computations are expressed as loops with necessary reductions inserted accordingly.
Arithmetic and bitwise logic operations on \texttt{slice} are further lowered into loops of operations on \texttt{row} agnostic to the backend, e.g., \texttt{row\_add} and \texttt{row\_and}.
A row operation's operand that the host produces and broadcasts across all PUD SIMD lanes need not be stored in DRAM rows. For each bit of this operand, \ToolName{} generates two simpler operations on the other operand alone for the host to select at run time, one for a zero and one for a one.
Conversely, intermediate results exchanged between chained offloaded operators can share reserved rows, binding the producing operator's result and the consumer's operand, which enables in-place reads and avoids a round trip to the host.

For the logic-level lowering, we present an \emph{architecture-aware logic optimization and code generator} (\autoref{sec:logicopt}) that constructs Boolean logic networks from these row-level operations. The logic-level optimization uses equality saturation to explore equivalent implementations over the target's instruction types, and greedy or ILP-based extraction methods to identify the optimal instruction sequence for the target architecture, e.g., a sequence of \texttt{AP} and \texttt{AAP} (\autoref{sec:background:PUD}).

We then map operations onto DRAM banks, treated as processing units (PUs). An ILP solver (\autoref{sec:ilp}) produces a mapping balancing cross-bank parallelism against synchronization overhead.
In the final code generation step, \ToolName{} lowers the row-level operations into \texttt{pud}-dialect instruction sequences, annotating \texttt{row}-typed operands with global word-line addresses and unrolling loops. The resulting code forms a dataflow graph dispatched across DRAM banks, driving correct instruction ordering and execution.

To evaluate our work, we further lower the \texttt{pud} IR to LLVM IR, replacing \texttt{pud} primitives with inline assembly encoding custom PUD opcodes. These are assembled into the compiled binary and decoded and executed natively by our extended gem5~\cite{oliveira2024mimdram,gem5} ISA decoder, alongside the surrounding CPU code, to simulate the full CPU-PUD system.

\section{Data Representation and Bit-Slicing}
\label{sec:bits}

To bridge word-level arithmetic and bitwise execution models, we introduce the \texttt{bits} dialect as an IR between \texttt{linalg} and PUD backends. This dialect makes explicit how bit-planes are physically organized within DRAM subarrays, since different PUD execution models impose different layout requirements: bit-parallel approaches~\cite{leitersdorf2023aritpim} uses standard word-oriented layouts, whereas bit-serial ones require a \emph{vertical} layout. For instance, in the latter case, a \texttt{tensor<NxiW>} is transformed into a \texttt{bits.slice<WxN>}, where each row represents one bit position and each column represents one tensor element, forming SIMD lanes across elements within a DRAM subarray. %
The conversion between \texttt{tensor} and \texttt{slice} is made explicit through two structural operations: \texttt{bits.transpose} and \texttt{bits.assemble}. How these are realized depends on the target: they may lower to an explicit data movement routine on the host CPU or a dedicated transposition unit~\cite{hajinazar2021simdram,shin2024processing}, or be elided entirely when data is already stored in required form and no physical movement is needed. This separates layout semantics from physical realization, keeping the dialect clean and avoiding unnecessary data movement.
After conversion to \texttt{slice}, computation is organized on bit-planes and needs further lowering to match row-based PUD operations, for which the \texttt{bits} dialect defines a \texttt{row} type, modeling a single row of a slice, which serves as the abstraction for operands of low-level PUD primitives.

\begin{figure*}
    \centering
    \includegraphics[width=0.90\textwidth,trim={0.5cm 0 1.5cm 0}, clip=true]{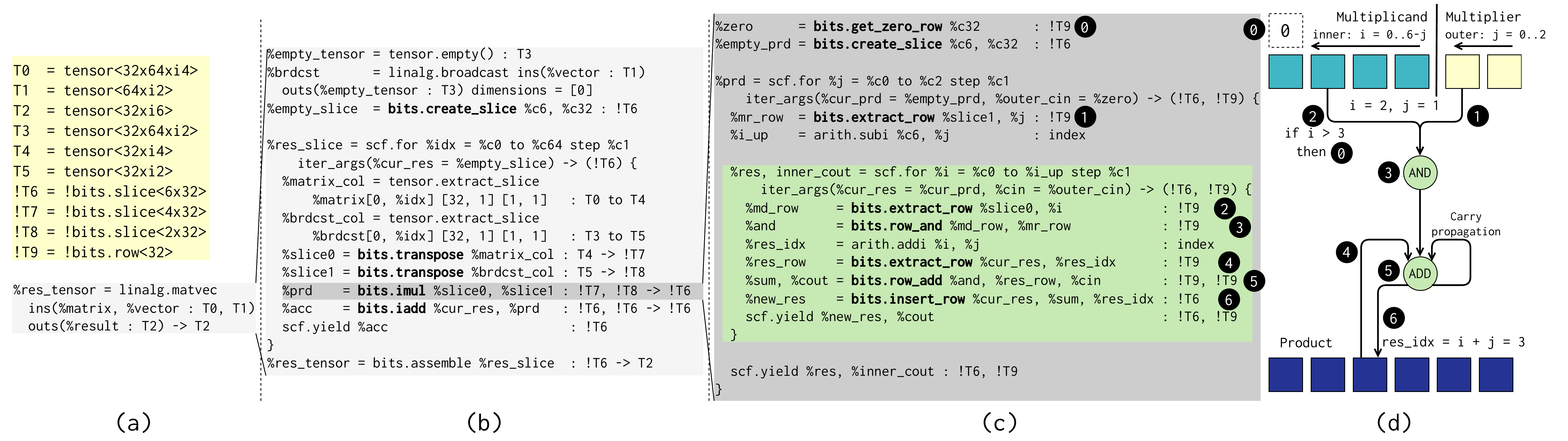}
    \caption{Progressive lowering of GEMV from \texttt{linalg} via slice-level to row-level \texttt{bits}.}
    \label{fig:bits}
\end{figure*}

Consider a \texttt{gemv} operation $\texttt{tensor<MxKxi4>} \times \texttt{tensor<Kxi2>}$ targeting a bit-serial substrate such as SIMDRAM. \autoref{fig:bits} illustrates the lowering from \texttt{linalg} to \texttt{bits} dialect and then from \texttt{slice} to \texttt{row}. This requires the compiler to (1) map the 2D matrix to 1D tensors with a loop over one dimension, and (2) enable efficient scheduling via a map-compute-reduce scheme, shown in \autoref{fig:bits}~(b). We first broadcast the vector operand along dimension \texttt{M} to match the matrix shape, then iterate over \texttt{K}: each iteration extracts a length-\texttt{M} sub-tensor from both operands and performs a bit-sliced multiplication followed by accumulation. The multiplication operands (\texttt{bits.slice<4x64>} and \texttt{bits.slice<2x64>} for \texttt{bits.imul}) are obtained by transposing the extracted tensors \texttt{tensor<64xi4>} and \texttt{tensor<64xi2>} on the CPU. Accumulation, as a reduction over \texttt{K}, can optionally be deferred until after the multiplication loop, letting all iterations run in parallel since there are no cross-iteration dependencies, though this trades bank-level parallelism against inter-bank data movement for reduction, a decision we discuss in \autoref{sec:ilp}.

After lowering from 2D tensors to bit-slices, the computation is further transformed to operate on bit-rows, i.e., DRAM wordlines. We implement an MLIR pass with predefined patterns that rewrite operations on \texttt{bits.slice<WxN>} into a loop with \texttt{W} iterations over \texttt{bits.row<N>}. For extensibility, our framework also allows users to define custom rewriting rules for structuring computation on bit-rows, for example to meet different computation requirements.
\autoref{fig:bits}(c) shows our rewrite of \texttt{bits.imul} into row-level operations like \texttt{bits.row\_and} and \texttt{bits.row\_add}, with key steps visualized in \autoref{fig:bits}(d). In this example, slice-level multiplication is lowered by iterating over multiplier bit positions. At each position, a single inner loop runs from the current shift offset through the most significant result row, generating row-wise partial products via \texttt{bits.row\_and} and accumulating them into shifted result rows using \texttt{bits.row\_add} with carry chaining. As the multiplicand rows are exhausted, the loop naturally transitions to pure carry propagation through the remaining higher result rows. The carry produced at the end of each outer iteration is forwarded as the initial carry of the next, ensuring correct chaining across multiplier bit positions.
In this way, \texttt{bits.slice<WxN>} multiplication is decomposed into row-wise $\texttt{extract} \rightarrow \texttt{and} \rightarrow \texttt{add} \rightarrow \texttt{insert}$ over \texttt{bits.row<N>} primitives, hence preserving arithmetic semantics in representing integer multiplication as bit-shifted additions in the bit-sliced manner during lowering.
Similarly, the slice-level \texttt{bits.iadd} on \texttt{bits.slice<WxN>} is lowered into \texttt{W} iterations of row-wise $\texttt{extract} \rightarrow \texttt{add} \rightarrow \texttt{insert}$ from least significant bit (LSB) to most significant bit (MSB), with correct carry propagation across iterations.

\section{Operation and Data Mapping}
\label{sec:ilp}

\begin{figure}[t]
    \centering
    \includegraphics[width=0.9\linewidth, trim= 1.3cm 0 1cm 0, clip]{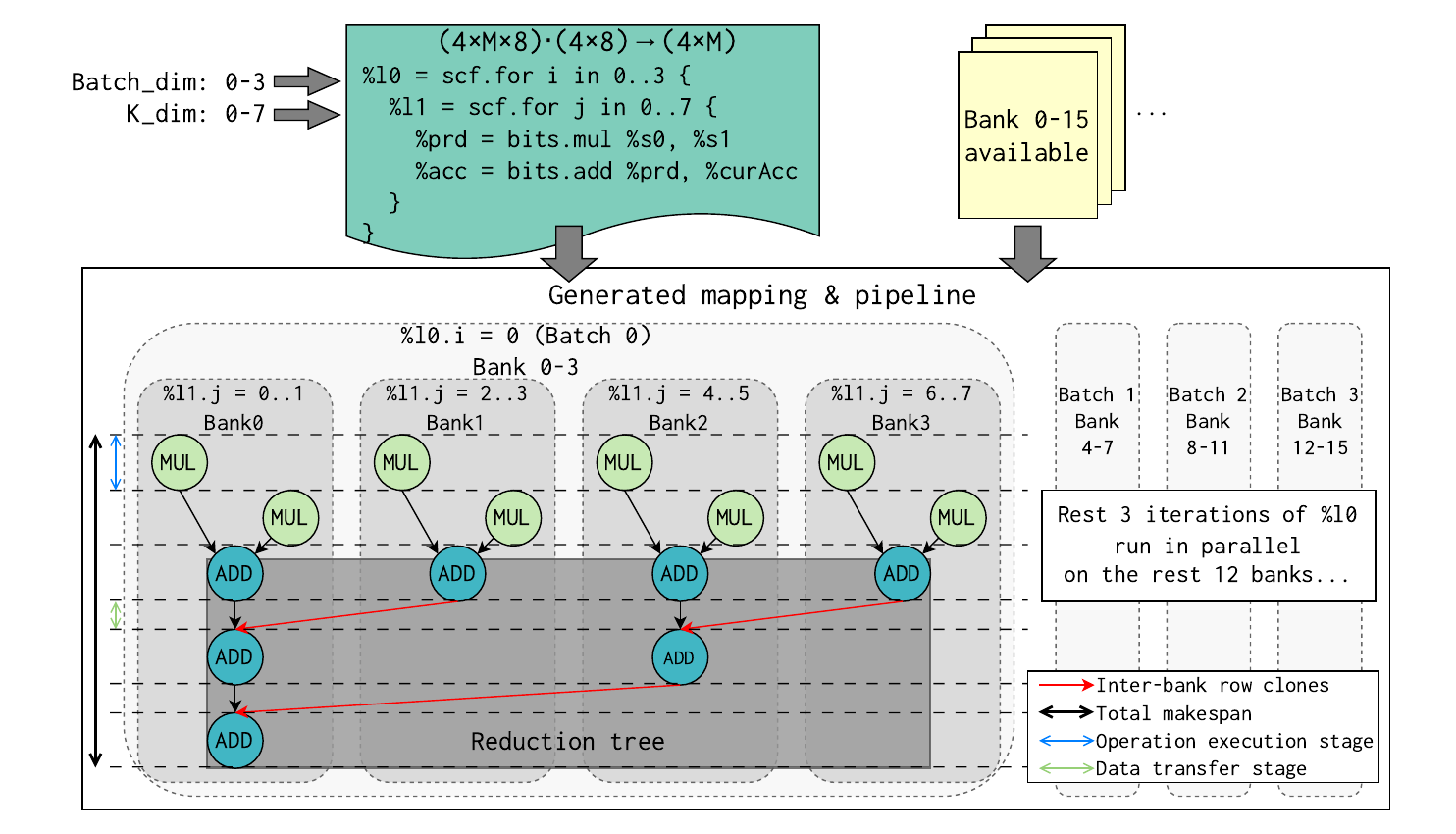}
    \caption{Ops-to-bank mapping, pipelining of batched GEMV.}
    \label{fig:ilp}
\end{figure}

To execute the batched GEMV in \autoref{fig:ilp} on multiple banks, the bank assignment and the schedule must be defined for each operation.
These decisions are inherently coupled, i.e., distribution on different banks increases parallelism but leads to inter-bank data transfer. We formulate this joint optimization as an ILP that minimizes the \textit{makespan}.

We map the arithmetic operations of \texttt{bits} dialect, i.e., those lowered to a fixed sequence of memory commands, excluding structural ones such as \texttt{bits.extract\_row}. We consider slice-level instead of row-level operations since all row-level operations derived from a single slice-level operation must execute on the same bank. Each operation to be mapped is associated with a latency estimated from the number of commands generated by the logic optimizer (see \autoref{sec:logicopt}) together with the iteration count given by the bitwidth.
For example, adding two $n$-bit integers costs $8n{+}1$ \texttt{AP/AAP} commands in SIMDRAM~\cite{hajinazar2021simdram} and $6n{+}1$ in PRADA~\cite{shin2024processing} (\autoref{fig:motiv}). The ratio varies from operation to operation, so a schedule priced for one substrate is not the one the other wants. We take one command as the unit of time.
For each operation, the ILP decides the bank assignment and the start time, while finish time follows from the operation's latency. Constraints enforce that each operation is assigned to exactly one bank and that data dependencies are respected.

Data placement introduces the first key constraint: values reside on the bank that produces them. If a consumer is placed on the same bank, execution proceeds without inter-bank transfer overhead; otherwise the value is copied one bit-plane row at a time, at a per-row cost that depends on whether the two banks share a channel. This is modeled via per-edge binary variables that select the applicable cost from the two operations' bank and channel assignments, and that add zero delay when they share a bank. Second, a bank is a sequential resource with one exception: as a bank row holds many elements, independent operations of the same kind and bitwidth can share a row and execute as one row-wide operation. Ordering variables serialize all other pairs, activated only when both operations share a bank, letting the solver trade off locality against parallelism. Third, transfers also consume bank resources: a transfer occupies both its endpoint banks for its duration, so we enforce non-overlap against other operations or transfers sharing an endpoint. An elaboration of our ILP formulation can be found in \autoref{tab:ilp}.

A loop body enters the ILP as individual nodes only when it carries neither a reduction nor a nested loop; every other loop nest becomes a single node with precomputed latency, as lifting one would grow the model past what the solver handles. For reduction loops, the precomputed latency is obtained by selecting the parallelism level that minimizes computation plus reduction overhead, used directly as the node cost. This level is not chosen in isolation: banks the reduction does not claim are taken by concurrent instances of itself from the enclosing parallel loops, which credit its local work but not its tree, whose transfers contend for the same channel buses. The two loops are therefore searched jointly, over how the banks are split between them.

Finally, we adopt two simplifications. First, the ILP prices a transfer by the location of its two banks but enforces only bank-level exclusion; contention on the channel data bus is modeled outside the ILP, in the closed-form estimator that prices reduction trees. Second, the loop nests priced in aggregate are atomic nodes, which limits global optimality.
The solving time is bounded by a configurable limit (300\,s by default): when it is reached, the best incumbent is used, and only if none exists do we fall back to a round-robin assignment.

\begin{table}[t]
\scriptsize
\centering
\caption{ILP formulation for operation-to-bank mapping.}
\label{tab:ilp}
\setlength{\tabcolsep}{4pt}
\renewcommand{\arraystretch}{1.1}
\begin{tabularx}{\columnwidth}{@{}>{\raggedright\arraybackslash}p{0.25\columnwidth} X@{}}
\toprule
\textbf{Symbol} & \textbf{Definition} \\
\midrule
\multicolumn{2}{@{}l}{\textbf{Sets \& Parameters}} \\
$\mathcal{N}=\{1,\dots,N\}$ & DAG nodes (ops / loops) \\
$\mathcal{B}=\{0,\dots,B{-}1\}$ & memory banks, partitioned into channels $\mathcal{K}$ \\
$\mathcal{E}\subseteq\mathcal{N}^2$; $\mathcal{C}\subseteq\mathcal{N}$ & DAG edges (dependencies); lifted loop nodes \\
$d_i \in \mathbb{Z}_{\geq 0}$ & Execution latency of node $i$ \\
$w_i, v_i$ & bit-plane rows and elements of its result \\
$\gamma^{\mathrm{in}}, \gamma^{\mathrm{x}}$ & Clone cost per row within / across channels \\
$W$; $M$ & bank-row capacity; Big-$M$ constant \\
\midrule
\multicolumn{2}{@{}l}{\textbf{Decision Variables}} \\
$x_{i,b}\in\{0,1\}$ & 1 iff node $i$ assigned to bank $b$ \\
$s_i, e_i \in\mathbb{Z}_{\geq 0}$ & Start, end time of node $i$ \\
$\mathit{cl}_{ij}\in\{0,1\}$ & 1 iff edge$_{ij}$ requires cross-bank clone \\
$q_{ijb}, q^{\mathrm{k}}_{ijk}\in\{0,1\}$ & Linearization aux.: $q_{ijb}=x_{i,b}\cdot x_{j,b}$, likewise per channel $k$ \\
$\sigma_{ij}$; $\tau_{ij}\in\{0,1\}$ & 1 iff nodes $i,j$ share a bank; share a channel (derived) \\
$o_{ij}$; $\pi_{ij}\in\{0,1\}$ & Ordering auxiliary for independent pair; 1 iff packed in one row \\
\midrule
\multicolumn{2}{@{}l}{\textbf{Constraints}} \\
C1; C2 & $\displaystyle\sum_{b\in\mathcal{B}} x_{i,b}=1; \; e_i=s_i + d_i
    \quad \forall i\in\mathcal{N}$ \\[2pt]
C3 & $q_{ijb} \leq x_{i,b},\quad q_{ijb} \leq x_{j,b},$\newline
   \hspace*{1em}$q_{ijb} \geq x_{i,b}+x_{j,b}-1
    \quad \forall (i,j)\in\mathcal{E}\cup\mathcal{P},\;\forall b\in\mathcal{B}$ \\[2pt]
C4 & $\sigma_{ij} = \displaystyle\sum_{b\in\mathcal{B}} q_{ijb},\quad
    \tau_{ij} = \displaystyle\sum_{k\in\mathcal{K}} q^{\mathrm{k}}_{ijk}
    \quad \forall (i,j)\in\mathcal{E}\cup\mathcal{P}$ \\[2pt]
C5; C6 & $\mathit{cl}_{ij} + \sigma_{ij} = 1;\; s_j \geq e_i + \gamma_{ij}
    \quad \forall (i,j)\in\mathcal{E}$\newline
   \hspace*{1em}$\gamma_{ij}=w_i\,[\,\gamma^{\mathrm{in}}(\tau_{ij}-\sigma_{ij})
    +\gamma^{\mathrm{x}}(1-\tau_{ij})\,]$ \\[2pt]
C7; C8 & $s_j \geq e_i - M\,(1-\sigma_{ij}+\pi_{ij}+o_{ij})$\newline
   \hspace*{1em}$s_i \geq e_j - M\,(1-\sigma_{ij}+\pi_{ij}+1-o_{ij})$\newline
   \hspace*{1em}$\pi_{ij}\leq\sigma_{ij},\;\; |s_i-s_j|\leq M\,(1-\pi_{ij}),\;\;
    v_i+\textstyle\sum_{j} v_j\,\pi_{ij} \leq W
    \quad \forall (i,j)\in\mathcal{P}$ \\[2pt]
C9 & $T \geq e_i
    \quad \forall i\in\mathcal{N}\setminus\mathcal{C}$ \\
\midrule
\textbf{Objective} & min (makespan) $T$ \\
\bottomrule
\end{tabularx}
{\scriptsize
$\mathcal{P}=\{(i,j)\mid i<j,\,(i,j)\notin\mathcal{E}^*,\,(j,i)\notin\mathcal{E}^*,\,
i\notin\mathcal{C},\,j\notin\mathcal{C}\}$: pairs with no dependency in either direction
($\mathcal{E}^*$ = transitive closure of $\mathcal{E}$). $\pi_{ij}$ exists only for pairs of equal kind and bitwidth. A clone excludes other work on its two endpoint banks over
$[e_i, e_i{+}\gamma_{ij}]$ by the big-$M$ pattern of C7, unless two clones share both
endpoints and one of them serves both.}
\end{table}

\autoref{fig:ilp} shows a batched GEMV (i.e., matrix shape \texttt{MxK}, vector length \texttt{K}, with $\texttt{K} = 8$ and batch size $4$) together with one mapping and schedule when 16 compute-capable banks are available.
The 16 banks are partitioned into four groups to enable full parallelism across the batch dimension. Each GEMV is assigned four banks to unroll the inner loop with factor 2. Each bank sequentially performs two multiplications and accumulates their results, followed by an inter-bank reduction implemented as an upside-down binary tree of two levels. Three inter-bank data copies are scheduled per GEMV, the two of the first level occurring in parallel. Although all banks follow the same pipeline in this example, the mapping pass supports assigning independent pipelines per bank. A global pipeline ensures that memory commands are issued to the correct bank at the correct time.

The reduction tree follows the memory hierarchy rather than the flat bank list: partial sums are halved within each channel first, on that channel's own data bus and in parallel across channels, and only the survivors are combined across channels. Pairing banks arbitrarily instead costs far more cross-channel traffic for the same result. The mapping pass not only constructs such a tree for efficient reduction, but also supports optional bitwidth optimization.
By default the accumulation wraps at the width the input program asks for; optionally, a sum of $n$ $b$-bit values is carried at the $b+\lceil\log_2 n\rceil$ bits it needs. A bit is added only when the count of combined values crosses a power of two: that addition keeps its carry-out as a new MSB and costs one command more, so growth is logarithmic in the number of values combined, not one bit per addition.
In \autoref{fig:ilp}, eight 8-bit products give slices of 9 rows after the first level and an 11-bit result at the root, and the transfers of partial sums are priced at the width the sum has reached.

Data mapping follows once operations are assigned to banks. A DRAM address is a tuple of channel, rank, bank, subarray, and row. Each \texttt{slice} is assigned contiguous rows within one subarray. An operand is assigned such a space if (1) it is obtained via \texttt{transpose} from a tensor, or (2) it is produced by an operation mapped to a different bank. Each result is always assigned one.

\section{Architecture-Aware Code Generation}
\label{sec:logicopt}

\begin{figure*}
    \centering
    \includegraphics[width=\textwidth]{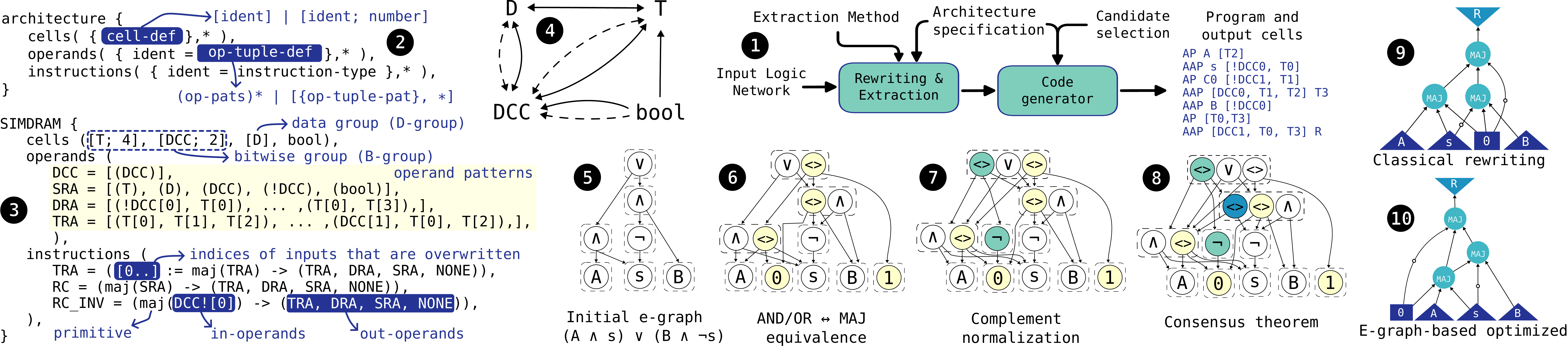}
    \caption{Overview of our code generator. The input is a circuit represented as a logic network; the output is an optimized command sequence with implicit row allocation.}
    \label{fig:logicopt}
\end{figure*}

A central challenge in mapping Boolean functions to PUD instructions (memory commands) is sharing optimizations across architectures. Prior code generators~\cite{chakraborti2014bdd,guo2025pimsynth,hajinazar2021simdram,marranghello2015sop,soeken2016mig,fayyazi2020hipe} each targeted a single architecture, and optimize the network for size and depth before, and independently of, row allocation (\autoref{sec:motiv}). We present a code generation engine (\autoref{fig:logicopt}) that retargets across PUD architectures by updating an architecture description file, without rewriting optimization passes.
In \ding{182}, the inputs are either an MLIR-lowered network or a standalone logic description, together with an architectural specification encoding PUD instruction constraints and costs. The engine performs e-graph-based rewriting and extraction, followed by candidate selection, to produce an optimized command sequence with implicit row allocation.

\subsection{Input Representation}
The input to the logic optimizer is a Boolean network: a DAG whose nodes are logic gates (\texttt{AND}, \texttt{OR}, \texttt{MAJ}, \texttt{XOR}) and whose edges carry a single-bit signal, optionally inverted. The optimizer accepts networks from two entry points: existing logic descriptions in BLIF, Verilog, or k-LUT format, imported via the mockturtle C++ library~\cite{EPFLLibraries} (providing standard types such as AND-Inverter Graph (AIG) and Majority-Inverter Graph (MIG)); or the \texttt{bits} dialect, whose operations (e.g., \texttt{bits.row\_add}, \texttt{bits.row\_mux}, etc., each producing a one-bit SSA value) are converted into a standard network type via the MLIR pipeline (lowering \texttt{bits.slice} to \texttt{bits.row}). For example, \texttt{R = torch.where(s, A, B)} lowers to \texttt{\%R = bits.row\_mux \%B, \%A, \%s}, corresponding to the logic network $(A \land s) \lor (B \land \lnot s)$ and represented in the initial e-graph as shown in \autoref{fig:logicopt}~(\ding{186}). Either way, the network is handed to the e-graph rewriter, which explores equivalent implementations over the target's instruction types.

\subsection{Architectural Constraints and Costs}
\label{sec:arch_def}
We model each PUD target through a structured architecture specification capturing three properties that jointly determine compilation cost: available cell types, operand placement rules, and instruction semantics.
\autoref{fig:logicopt} (\ding{183}) illustrates the specification syntax %
mapping to a hierarchy of \textit{structs} and \textit{enums}: \texttt{cells} carry a name and optional capacity, and \texttt{operands} group cells together with an optional inversion flag and \emph{operand patterns}, which cells and polarities are legal at a given instruction input or output position. Finally, \texttt{instructions} hold the primitive as an enum variant (\texttt{maj}, etc.), a list of input operand pattern sets, a list of output operand pattern sets, and a set of destructive input positions. %

Architectural constraints arise first from the operand pattern system: each instruction type statically restricts which cell types, and with which polarity (inverted or not), may occupy each operand slot. For example, \autoref{fig:logicopt} (\ding{184}) presents the concrete specification for SIMDRAM~\cite{hajinazar2021simdram}, where the highlighted \emph{operand patterns} define which combination of rows are allowed for \texttt{MRA}.
The second source of constraints is the \emph{copy graph} (\autoref{fig:logicopt}, \ding{185}), a directed graph whose nodes are cell-type patterns and whose edges encode the cheapest non-destructive copy operations between them, annotated with their cost and whether the copy is inverting (dashed edges). For SIMDRAM, the cell types are DCC, data (D), bitwise (B), and constant \texttt{0/1} rows; the copy graph \ding{185} encodes how signals may move between them.

The copy graph is automatically derived from the instruction types at startup by symbolic evaluation, so the user never specifies it manually.
During code generation, whenever a signal must be moved to satisfy an operand constraint, shortest-path search (Dijkstra) over the copy graph determines the cheapest legal sequence of copy instructions and this overhead is what distinguishes architectures with flexible addressing such as PRADA~\cite{shin2024processing} from those with rigid row-allocation restrictions such as SIMDRAM \cite{hajinazar2021simdram} or neighboring subarrays restrictions such as FCDRAM~\cite{yuksel2024functionally}. This copy overhead feeds into the cost function, which maps each instruction type to a non-negative integer; the total program cost is the sum of all instruction costs, including copies.

\noindent\textbf{Scope of the description.} A target is expressible if it satisfies two conditions: (i) each primitive must operate on whole rows to evaluate an \texttt{AND}, \texttt{OR}, \texttt{MAJ}$_k$, or \texttt{XOR} gate, with optional inversion of inputs and output, or to write a constant; (ii) the target's constraints must be expressible in terms of the cells and polarities available to each operand and the inputs overwritten by each primitive (\autoref{sec:arch_def}). This covers designs with dedicated compute rows (SIMDRAM, Ambit~\cite{seshadri2017ambit}), designs with flexible row roles (PRADA), and unmodified COTS DDR4 (FCDRAM), each described in under 20 lines. \autoref{tab:retarget} lists what a new target requires in \ToolName{}, compared with PIMsynth.

\begin{table}[t]
\scriptsize
\centering
\caption{Retargeting effort against PIMsynth.}
\label{tab:retarget}
\setlength{\tabcolsep}{3pt}
\begin{tabular*}{\columnwidth}{@{\extracolsep{\fill}}lll@{}}
\toprule
\textbf{Component} & \textbf{PIMsynth~\cite{guo2025pimsynth}} & \textbf{\ToolName{}} \\
\midrule
Front end, bit-slicing, row-level lowering & reused; RTL per kernel & reused \\
Tensor-level mapping and scheduling & not provided & reused \\
Rewrite rules & none (ABC mapping) & subset selected \\
Command encoding & per command, per mode & reused if shared \\
Architecture description, cost weights & cell library; unit cost & $<$20 lines \\
Copy graph, extraction costs & hand-written passes & derived \\
Command sequence and cost per operation & hand-written per gate & derived \\
Reserved-row address map & hand-written & derived \\
\bottomrule
\end{tabular*}%

\end{table}

\subsection{Equality Saturation Step}
In \ToolName{}, we use the \texttt{egg} library~\cite{willsey2021egg}, as no open-source e-graph implementation integrated with MLIR was available at the time of this work. Applying equality saturation (\autoref{sec:background:egraph}) to our setting, the initial logic network for a given Boolean function is seeded in the initial e-graph (\autoref{fig:logicopt}, \ding{186}): each gate and signal becomes an e-class (dashed rectangles). Applying the rewriting rules (a set of which is listed in~\cite{amaru2014majority,chu2019structural}) grows the equivalence classes to include several derived alternatives (steps \ding{187}, \ding{188}, and \ding{189}). Importantly, the rule set is filtered per architecture: if an architecture only has \texttt{MAJ} primitives, only \texttt{MAJ}-related rules are activated; if it has \texttt{MAJ} and \texttt{XOR}, like in~\cite{zhou2022flexidram,soleimani2025wisedram}, cross-rules such as ``\texttt{MAJ} distributes over \texttt{XOR}'' are also included~\cite{chu2019structural}.

\subsection{Extraction: Picking the Best Candidate}
\label{sec:extraction}
Equality saturation decouples rewriting from selection: all equivalent networks are accumulated in an e-graph, and extraction applies architecture-specific costs and placement constraints globally once rewriting is complete, choosing the implementation that minimizes instruction count for the target rather than gate count in isolation.
After rewriting, we must pick a concrete representation of the logic network in the e-graph for code generation. This extraction problem is where architecture-specific costs and constraints are applied.%
We devise three extraction approaches:

\noindent\textbf{Code Generation Guided Extraction.} The simplest approach is to use the actual code generator as the cost oracle.
Starting with the root e-node, we iterate over the representatives in the e-node and generate code for the partial logic network obtained from it. We greedily pick the lowest actual program cost and recurse over the children e-nodes.
Because the cost is the real generator output, each local choice is scored on true allocation overhead; but the greedy, per-cone commitment means the assembled program can still be globally suboptimal.
This approach can be slow because running code generation for every candidate is expensive.

\noindent\textbf{Local Cost Function Extraction.} This approach rewrites the e-graph so that architecture costs become local, per-node costs. The PUD-specific difficulty is \textit{asymmetric polarity costs}: some naturally deliver an inverted result, so obtaining the opposite polarity requires an extra copy. We capture this by splitting each e-class into two \emph{access e-classes} (inverted and non-inverted) where accessing the polarity an instruction produces natively is free, and the other pays a copy cost estimated from the architecture's copy graph. Every e-node then carries a local cost (instruction cost for gates, copy cost for access nodes), enabling standard extraction with \texttt{egg}'s ILP~\cite{willsey2021egg} to find the cost-optimal representatives under this model or use a cheaper greedy variant that might ignore shared common-subexpressions.

\noindent\textbf{Compile-off Extraction.} Local cost extraction is optimal only for the estimated cost: the copy overhead it charges is an approximation from the copy graph, so the single network it returns is not guaranteed to compile to the cheapest program. Rather than trust one extraction, we generate several candidate networks cheaply and let the real code generator decide among whole programs from either cost-perturbed greedy extraction (adding deterministic per-node jitter to diversify the structures produced) or ILP solutions. Each candidate is compiled in full, and we keep the one with the lowest actual program cost. Unlike code-generation-guided extraction, which commits greedily cone-by-cone, compile-off compares complete assembled networks, capturing cross-output sharing and the copies that allocation induces globally.

The result is a logic network (\autoref{fig:logicopt},~\ding{191}) tailored to the target architecture: it uses only available primitives, in a form that reduces copy overhead induced by the row allocation rules, with instruction cost minimized according to the target's cost function. \autoref{fig:logicopt}~(\ding{190}) shows the solution from state-of-the-art classical rewriting~\cite{EPFLLibraries}, versus the e-graph-based result in \ding{191}, reducing instruction count from 8 to 7 \texttt{AP/AAPs}. The resulting network is then fed into the code generator to produce the final \texttt{pud} dialect instruction sequence.

\section{Experimental Setup and Evaluation}
\label{sec:eval}
\noindent\textbf{Simulation infrastructure}. %
We evaluate \ToolName{} on a host-PUD system built on gem5-CIMD~\cite{kusnezoff2026gem5}, already validated against \cite{hajinazar2021simdram,oliveira2024mimdram}, simulating both HBM3-4Gb-x64 and DDR4-2400-x64 across three PUD architectures (\autoref{sec:background:PUD}): SIMDRAM~\cite{hajinazar2021simdram}, PRADA~\cite{shin2024processing}, and FCDRAM~\cite{yuksel2024functionally}.
Each target is added as an architecture description (\autoref{sec:arch_def}, \autoref{tab:retarget}); no optimization pass is modified. We use two execution modes to capture the effects relevant to each experiment. For \autoref{tab:ann}, where host phases contribute to runtime, we use gem5-CIMD's full-system mode: the x86 core issues each row command as an instruction. For the schedule comparisons in \autoref{fig:cmp_optiPIM}--\autoref{fig:interaction}, we instead execute the compiled schedules of both \ToolName{} and OptiPIM directly at the memory controllers. %
This excludes host-side effects such as code placement and accommodates OptiPIM's output, which consists of command streams rather than host programs. %

\noindent\textbf{Baselines.} For tensor-level optimizations, we compare to OptiPIM.\footnote{https://github.com/ETHZ-DYNAMO/OptiPIM/ -- commit 9be8f9a.}
OptiPIM's code generator and simulator make several approximations that would make a direct comparison to our gem5 model unfair. We correct them by: (i) porting OptiPIM's code generation from Ramulator to gem5 and driving it through the identical DRAM controller, (ii) modeling the $n$-bit \texttt{imul} cost for SIMDRAM as $\approx 13n^2$ operations (versus OptiPIM's $7n^2{+}1$ operations, which omits the RowClone operations)
and (iii) computing its ILP cost bounds in floating point, as 64-bit integers overflow beyond $2^{32}$ MACs.
By default OptiPIM simulates a single representative bank and assumes the other $N{-}1$ execute in perfect lockstep; our gem5 model captures realistic bank parallelism with $t_{RRD}$ pipelining under $t_{FAW}$ window. For logic-level optimization, we compare against PIMsynth~\cite{guo2025pimsynth}, which among our targets supports only SIMDRAM, and against a baseline built on mockturtle~\cite{EPFLLibraries}. Mockturtle is an open-source logic synthesis library that implements the cut-based rewriting and algebraic MIG optimizations~\cite{amaru2014majority} used in standard ASIC/FPGA flows, the same class of optimizers that SIMDRAM relies on. Both baselines minimize network size and depth but are unaware of the target’s row allocation and copy costs.

\noindent\textbf{Faithfulness to SIMDRAM.} We reproduced SIMDRAM's row allocation and $\mu$Program generation from their description~\cite{hajinazar2021simdram}. Applied literally, the procedure yields no correct $\mu$Program. We therefore compare against an \emph{optimal reconstruction} of SIMDRAM, an exhaustive allocation search under its command set. This matches SIMDRAM's published counts for add, sub, the comparisons, if-else, and max/min, though not for equal and ReLU. Since this reconstruction is optimal for a given MIG, the cost is determined by the MIG itself, which SIMDRAM selects without regard to allocation.

\subsection{Logic-Level Optimizations}
\begin{figure}[t]
    \centering
    \begin{subfigure}{\linewidth}
        \centering
        \includegraphics[width=\linewidth]{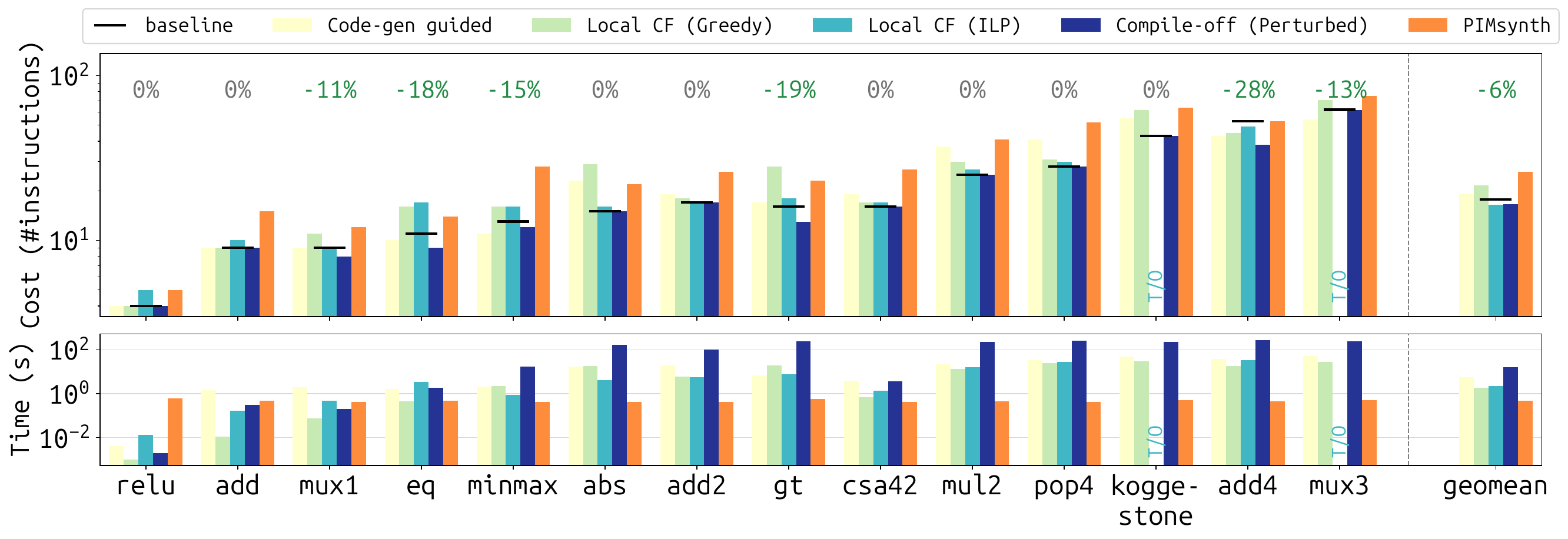}

        \caption{SIMDRAM}
        \label{fig:results:simdram}

    \end{subfigure}
    \begin{subfigure}{\linewidth}
        \centering
        \includegraphics[width=\linewidth]{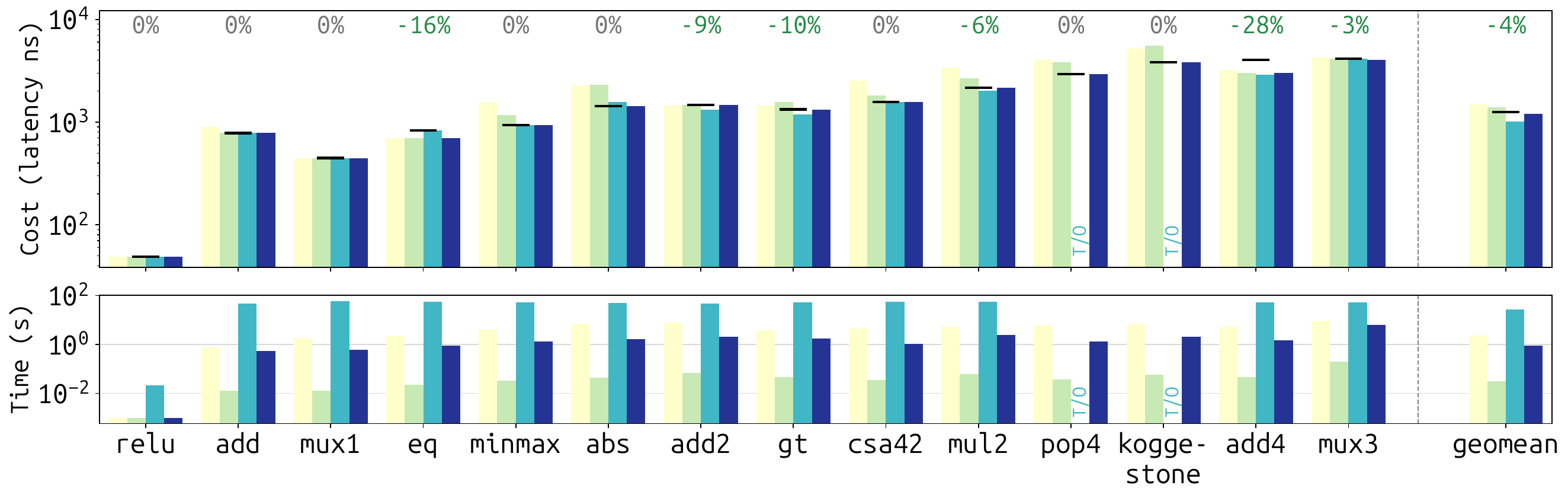}
        \caption{PRADA}
        \label{fig:results:prada}
    \end{subfigure}
    \begin{subfigure}{\linewidth}
        \centering
        \includegraphics[width=\linewidth]{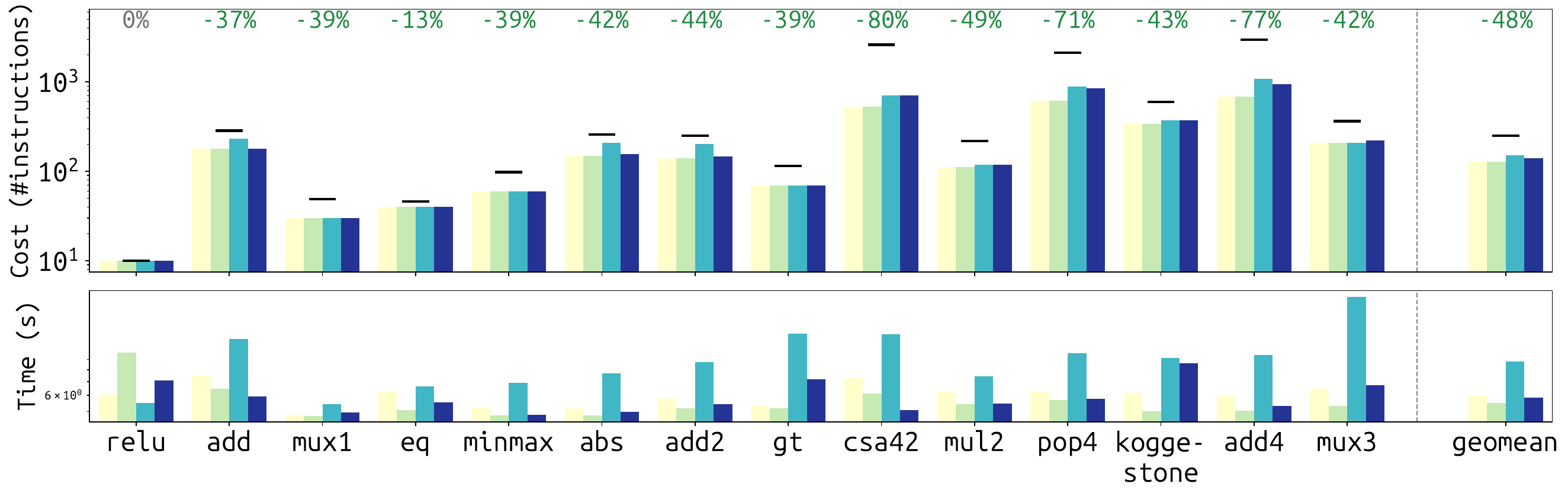}
        \caption{FCDRAM}
        \label{fig:results:fcdram}
    \end{subfigure}
    \caption{Program cost (instruction count or latency) and optimization time across extraction methods.}
    \label{fig:results:comparison}
\end{figure}

This section shows that feeding allocation cost back into candidate selection over an e-graph enables PUD synthesis to generate lower-cost programs than state-of-the-art methods. We retarget the same synthesis flow to SIMDRAM~\cite{hajinazar2021simdram}, PRADA~\cite{shin2024processing}, and FCDRAM~\cite{yuksel2024functionally}, reporting program quality (instruction count or latency) per extraction strategy and end-to-end optimization time (time the logic optimizer takes to generate instructions sequence), across 14 kernels, in \autoref{fig:results:comparison}.

The \emph{baseline} represents prior PUD synthesis approaches~\cite{hajinazar2021simdram,oliveira2024mimdram} that treat network optimization and gate-to-row allocation as separate, sequential phases. The baseline applies no e-graph rewriting, but the network is still mockturtle-preoptimized and then handed to the exact same list scheduler as every other strategy. On SIMDRAM, PIMsynth's geomean cost is 1.47$\times$ higher than the baseline. \autoref{fig:results:comparison} compares the three extraction methods from \autoref{sec:extraction} against the baseline; the cost reduction obtained by the best strategy per kernel is annotated on top of the bars.

The greedy and code-gen-guided extractors perform poorly compared to no rewriting in \autoref{fig:results:simdram} and \autoref{fig:results:prada}. ILP-based Local CF and perturbed compile-off are, on average, the best strategies across all architectures. ILP-based Local~CF is a strong method: it is the only one that beats the compile-off on PRADA, but it does not scale, timing out (300s) on the larger kernels. Compile-off matches or exceeds ILP quality without timing out, and never regresses below the baseline.

Against decoupled approaches, compile-off reduces cost by up to 25\%, while improving over no-rewriting by 48\% for FCDRAM and 4--6\% for SIMDRAM and PRADA. %
The mechanism is the same in every case: only compile-off evaluates the whole-program cost, so it accounts for the copies that destructive in-memory computation inserts when a sub-expression is shared across outputs, which is an overhead that code-gen guided and Local~CF extractors cannot see. Compile-off extraction compiles a small pool of candidate networks and keeps the cheapest; its compilation time ranges from seconds to a few minutes per kernel and scales with the number of candidates. The ILP-based extractors are a quality reference but time out on the larger kernels. %

\noindent\textbf{The effect is architecture-dependent.} PRADA makes the mechanism most visible: because a copy costs about twice a compute operation, a copy-blind optimizer that duplicates shared logic pays double, and the benefit of coupling appears directly in \emph{latency}. We use latency as PRADA's quality metric because its instructions have widely differing latencies; this also exercises the retargetability of the cost function. FCDRAM (\autoref{fig:results:fcdram}) is the opposite regime. Its un-rewritten networks are large, so \emph{every} strategy improves substantially ($0.52\times$--$0.60\times$) and the allocation-cost that distinguishes the strategies matters less.

\begin{figure}
    \centering
    \includegraphics[width=\linewidth]{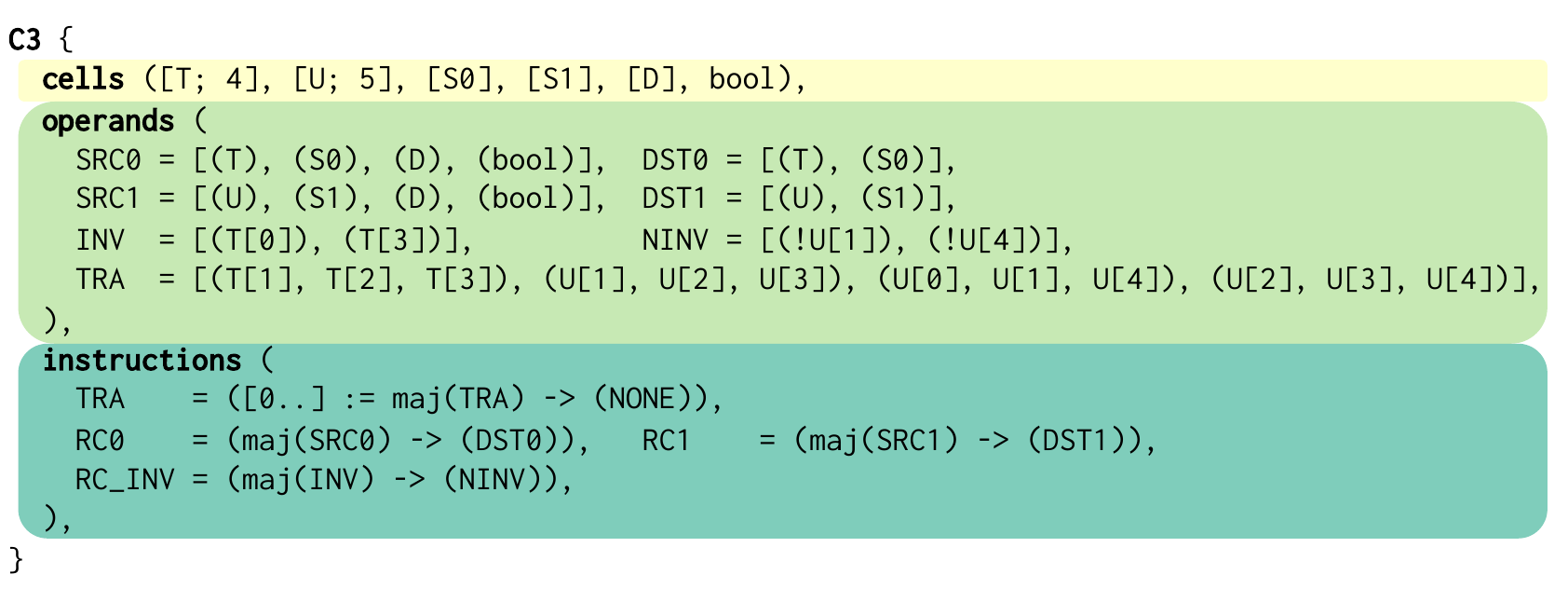}
    \caption{\Cthree{}~\cite{ji2026c3}'s primitive set as a \ToolName{} description.}
    \label{fig:c3}
\end{figure}

\noindent\textbf{Retargeting effort.} We retarget \ToolName{} to \Cthree{}~\cite{ji2026c3}, a recently proposed PUD substrate that was validated on commodity DDR4. It keeps only the primitives that proved reliable there, forcing every circuit through \texttt{MAJ3} and a negation that is a copy into the neighbouring subarray. \autoref{fig:c3} is its architecture description in \ToolName{}. From this alone, \ToolName{} automatically derives a full adder program of 16 commands: 10 copies, 2 inverting copies, 4 \texttt{MAJ3}. This command sequence matches the one presented in \Cthree{}~\cite{ji2026c3}, which was written by hand and validated on silicon. For retargeting, \ToolName{}'s front-end, tensor-level mapping, and rewrite rules are reused, while the copy graph, command sequences, and costs are derived from the supplied description. Compared to prior work, a target-specific flow like SIMDRAM builds its logic decomposition, row allocation, cost model, and $\mu$Programs around one primitive set, so every variant means rewriting them.
PIMsynth accepts different gate libraries, but it describes compute rows only by their number and assumes each provides negation, so \Cthree{}'s fixed row triples and cross-subarray negation are outside its description.
In \ToolName{}, a different row selection is an edit to the operand patterns (\autoref{fig:logicopt}, \ding{184}), after which every circuit is recompiled with no further input.

\noindent\textbf{Standard logic synthesis is under-optimized for PUD.} Prior PUD flows borrow logic optimization from ASIC/FPGA synthesis: SIMDRAM applies algebraic MIG rewriting~\cite{amaru2014majority}, and PIMsynth relies on ABC~\cite{guo2025pimsynth,brayton2010abc}. These tools minimize gate count and depth under a cost model in which fan-out is free and gate inputs survive evaluation. However, these minimization assumptions do not always hold in PUD. \autoref{fig:results:comparison} shows that our baseline and PIMSynth, which have networks with optimal number of gates, can therefore compile to \emph{more} commands (exemplified by \autoref{fig:logicopt}, \ding{190} vs.\ \ding{191}), so logic optimization must see the target's allocation and copy costs.

\subsection{Tensor-Level Optimizations}

\begin{figure}
    \centering
    \includegraphics[width=\linewidth]{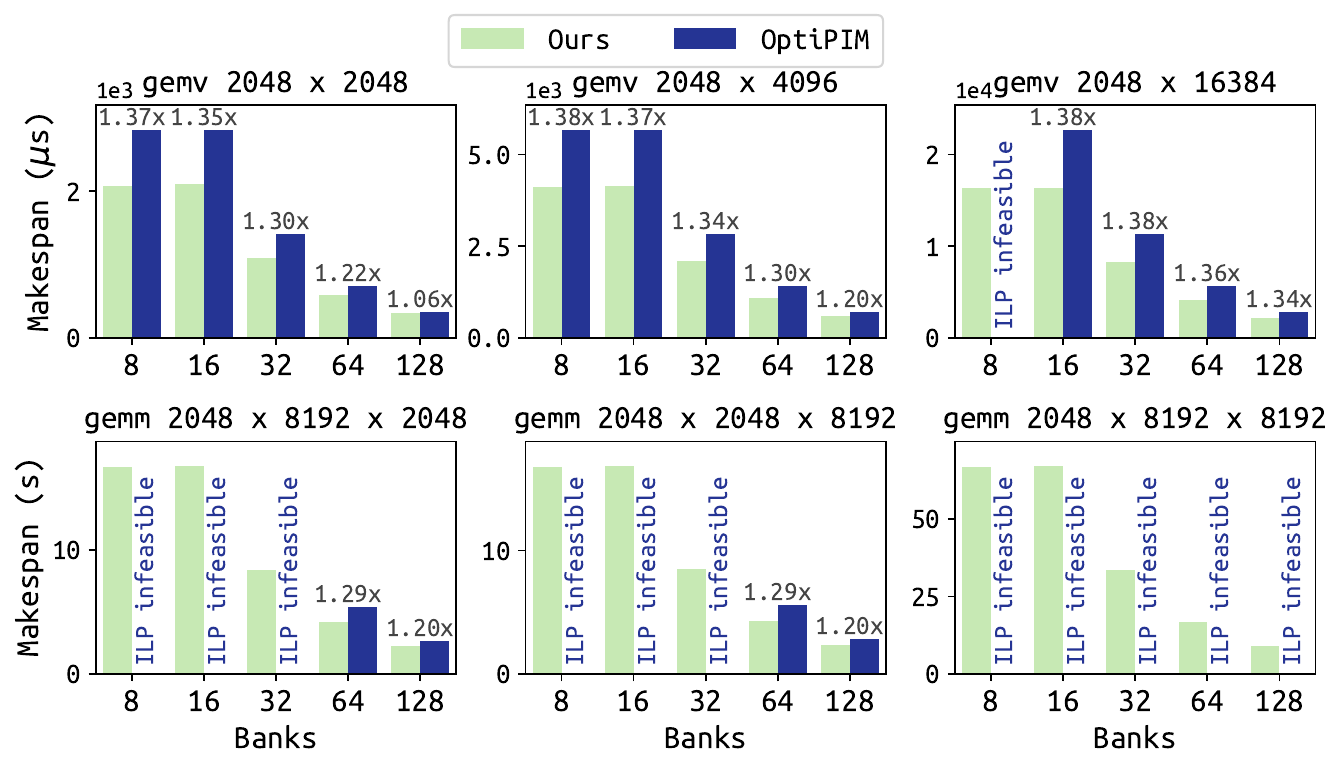}
    \caption{Makespan of ours vs. OptiPIM over \texttt{gemv} (top) and \texttt{gemm} (bottom), 8--128 banks on HBM3.}
    \label{fig:cmp_optiPIM}
\end{figure}

In this section, we show that our ILP-based mapping and scheduling is 6--38\% faster and, unlike OptiPIM~\cite{liu2025optipim}, produces a valid layout for every shape and bank count. We evaluate \ToolName{} against OptiPIM by sweeping \texttt{gemv} and \texttt{gemm} sizes across bank counts from 8 to 128, using \texttt{int8} inputs and \texttt{int32} accumulation. \autoref{fig:cmp_optiPIM} reports makespan under a fair, in-order execution model on an HBM3 substrate.

Our schedule is more efficient for two reasons. First, we perform the reduction with in-array copies rather than streaming partial sums through the HBM I/O (eliminating the 37\% of OptiPIM's command stream spent on readouts). This is the dominant cost at low bank counts, and it is why the gap is largest (up to 1.38$\times$) at 8 banks.
As bank count grows, both schedules become \texttt{ACT}-bound rather than readout-bound, and in-array reduction is increasingly hidden behind parallel work, and speedups reducing to 1.06--1.34$\times$ at 128 banks. This comparison favors OptiPIM, since its cost model charges only for \texttt{imul} and treats \texttt{add} as free, despite addition accounting for 13.4--15.2\% of the \texttt{gemv} command sequence. Costing addition, either through in-memory or by an additional logic outside of the memory, would only widen the gap in our favor.

Second, \ToolName{} exposes fine-grained, independent row operations that keep the per-bank activation pipeline full at $t_{RRD}$/$t_{FAW}$ constraints, whereas OptiPIM's coarse per-multiplication broadcast serializes reduction steps: bit-sliced \texttt{imul} and \texttt{add} become independent macro-ops the ILP can pipeline across banks without barriers, while OptiPIM issues operation-major broadcasts with a per-timestep barrier that holds the next reduction step until the current one finishes on all $N$ banks, leaving the pipeline underfilled.

Separately, OptiPIM formulates data layout as an ILP that places the entire $M\times K\times N$ iteration space onto the HBM at once, keeping each reduction fully resident with no path to spill or stream partial sums. Placeable volume is thus capped by the product of per-column, per-bank, and bank capacities ($M\cdot K\cdot N \le 2^{29}B$ for this HBM configuration, with $B$ the number of banks), which grows linearly with bank count $B$: large GEMMs at low bank counts exceed this bound, making the ILP infeasible, whereas \ToolName{} tiles the reduction temporally and maps every shape.

\begin{figure}
    \centering
    \includegraphics[width=\linewidth]{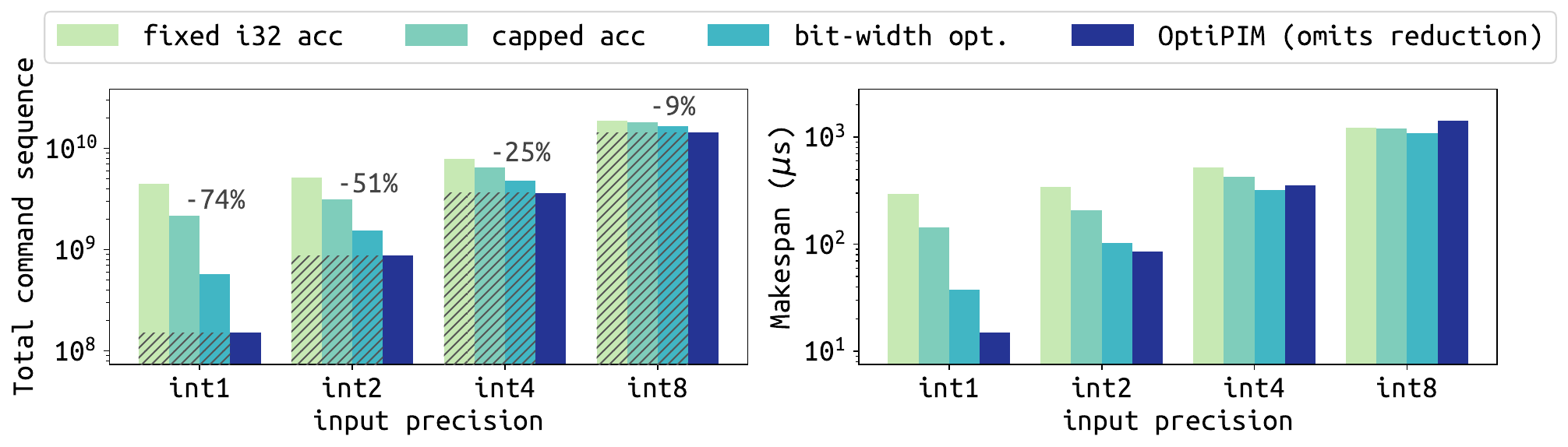}
    \caption{Command count (left) and 128-bank makespan (right) for \texttt{gemv} $2048\times8192$ across \texttt{int1}--\texttt{int8} input operands.}
    \label{fig:bitwidth_opt}
\end{figure}

\noindent\textbf{Bitwidth-aware reduction.} Our compiler is the first to introduce bitwidth analysis in the automation flow for PUD architectures. By identifying the minimum number of bits required to represent data at each point in a computation, we reduce the number of memory commands by  11--87\%  while maintaining accuracy and correctness. Fig.~\ref{fig:bitwidth_opt} sweeps input precision from \texttt{int1} to \texttt{int8} for a \texttt{gemv} ($2048\times8192$), reporting both the command count (left) and the makespan at 128 banks (right). Because operands are bit-serial, a naïve lowering that accumulates every partial sum into a \texttt{int32} accumulator pays a fixed per-add cost. Our reduction instead grows the adder width by one bit per adder tree level, so each addition costs exactly what its operands require.
The benefit is largest at low precision and shrinks as $b$ grows. At high precision, the multiply term ($\Theta(b^2)$) dominates and the fixed accumulator is already close to the working width, leaving little to trim, whereas in the quantized regime that makes PUD attractive (\texttt{int1}--\texttt{int2}) the reduction is the overwhelming cost and our compiler removes most of it.

A fixed-width baseline, which caps the accumulator at the final reduction width of $2b+\log_2 K$ bits (\emph{capped acc}), recovers part of this gap, but still pays the widest width on the many early tree levels where operands are narrow, so per-level widening reduces commands by a further 9--74\% over the fixed-width baseline. OptiPIM's advantage in this figure is an artifact of an incomplete cost model rather than a more efficient schedule. As OptiPIM treats the reduction as free, its bar presents the multiply-only floor. The hatched region on the \textcolor[HTML]{C7E9B4}{green} and \textcolor[HTML]{41b6c4}{teal} bars presents multiply floor for \ToolName{}, confirming that they have the same multiply cost. The gap between the \textcolor[HTML]{253494}{dark blue} and \textcolor[HTML]{41b6c4}{teal} bar (hatched vs. unhatched) is precisely the reduction OptiPIM omits, and that omission
is most consequential exactly where PUD is most useful, at low bitwidth.

\noindent\textbf{Compile time and scalability.} Single \texttt{int8} \texttt{gemv} and \texttt{gemm} operators compile end-to-end in 0.77--0.92\,s across $2^{22}$--$2^{32}$ MACs and 8--128 banks. The logic optimizer (\autoref{fig:results:comparison}) accounts for 80--92\% of this time and depends only on which row operations a kernel uses. Mapping takes 12\,ms regardless of tensor size and bank count and grows only with the number of operators. For 64 \texttt{gemv} layers on 128 banks, the ILP has 780k constraints and is solved to optimality in 11.4\,s. On 8 banks the solver reaches its 300\,s limit from 32 layers, and compilation continues with the best incumbent. The round-robin fallback was never needed.

\subsection{Interaction between Optimization Levels}

Prior PUD compilers either omit tensor-level optimization~\cite{peng2023chopper, guo2025pimsynth, de2025proteus} or do it by utilizing a hand-derived cost model~\cite{liu2025optipim}. \ToolName{} is the first that makes tensor-level mapping and scheduling decisions using per-primitive costs automatically supplied from the logic level and generates logic-level programs with tensor-level information such as bitwidth analysis. The value of this interaction between the two coupled optimization levels is quantified in this section.

\begin{figure}
    \centering
    \includegraphics[width=\linewidth]{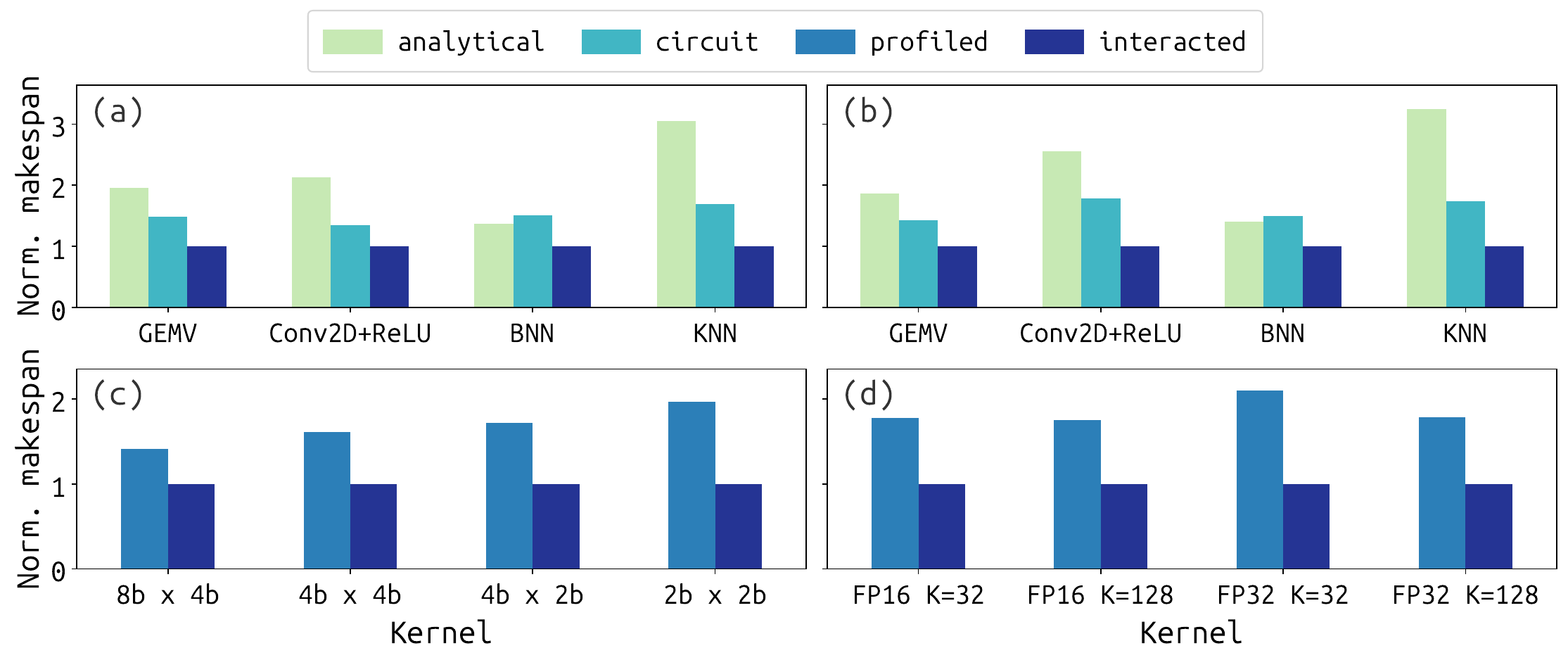}
    \caption{Makespan on 128-bank DDR4, normalized per kernel to the mapping that interacts with the logic level. (a)~SIMDRAM; (b)~PRADA; (c)~SIMDRAM: one \texttt{int8} \texttt{gemv} kernel with four approximate multipliers; (d)~SIMDRAM: four \texttt{FP} \texttt{gemv} kernels with one approximate multiplier.}
    \label{fig:interaction}
\end{figure}

\autoref{fig:interaction}~(a) and (b) compare mappings produced with two decoupled cost models against the one that interacts with the logic level. The \emph{analytical} model uses all information about the operation available at the tensor level such as bitwidth and complexity, while the \emph{circuit} model is architecture-specific and prices every operation by the number of gates in its logic network. Both models know the correct cost of inter-bank data transfer on DDR4. Across four kernels on SIMDRAM and PRADA, \emph{analytical} is $1.37$--$3.25\times$ slower than \emph{interacted}. \emph{circuit} is closer on three of the four kernels but still loses on every one, $1.34$--$1.78\times$. The reason is that the trade-off between data locality and parallelism for optimizing tensor-level loops can never be made correctly until the exact computation latency is known, which requires an accurate cost for each operation, derived from the compiled logic-level program.
With the transfer latency known, overestimating computation over-unrolls loops and makes execution transfer-bound, while underestimating it leaves banks unused and makes execution compute-bound.

A decoupled mapper can also be given measured costs, for instance profiled latency of each operation on the target. For a fixed lowering from one program to one substrate, the profiled cost model leads to the same scheduling as \ToolName{}'s. However, for each lowering one such profiling is necessary. Since \ToolName{} allows users to define their own arithmetic algorithms (\autoref{sec:bits}), \autoref{fig:interaction}~(c) changes the row-level implementation of \texttt{bits.imul} in an \texttt{int8} \texttt{gemv} ($2048\times128$) using four configurations of the broken-array multiplier~\cite{mahdiani2010bioinspired}, which keeps the top $A{\times}B$ bits and drops the partial products of the low bits, while (d) implements four \texttt{gemv} kernels' \texttt{bits.fmul} in an approximate manner proposed in~\cite{guo2025pimfp}. Compared to \ToolName{}, a decoupled mapper with the profiled costs of the exact \texttt{int8}, \texttt{FP16} and \texttt{FP32} multipliers generates mappings that are $1.42$--$2.1\times$ slower. In both panels the profiled costs are correct for the program the tensor level sees. The loss comes only from a change at the logic level that did not reach the mapper and the mapper then splits the cheaper computation across more banks than it is worth. To achieve an optimal mapping, a decoupled flow has to re-profile every operation, bitwidth, and substrate after each such change. In \ToolName{} this happens on every compilation automatically through the interaction between the two optimization levels.

\subsection{End-to-End Compilation: Vector Search}

\begin{table}[t]
\centering
\caption{Per-query latency (ms) and DRAM energy (mJ) of ANNS on SIFT1M; each row compiles one more part to PUD.}
\label{tab:ann}
\footnotesize
\setlength{\tabcolsep}{1.8pt}
\begin{tabular*}{\columnwidth}{@{}l@{\extracolsep{\fill}}rrrrrrrrr@{}}
\toprule
 & Proj. & Bcast & Scan & Select & Rerank & Total & Speedup & Energy & Saving \\
\midrule
CPU-only               & 0.01 & --   & 5.07 & 1.05 & 0.26 & 6.39 & --   & 39.0 & --   \\
Kernel offload         & 0.05 & 0.07 & 0.59 & 1.18 & 0.27 & 2.18 & 2.93$\times$ & 15.1 & 2.59$\times$ \\
+ Broadcast spec.      & 0.05 & --   & 0.46 & 1.17 & 0.27 & 1.98 & 3.23$\times$ & 13.4 & 2.90$\times$ \\
+ Kernel chaining      & 0.04 & --   & 0.46 & 0.35 & 0.25 & 1.10 & 5.84$\times$ &  8.1 & 4.79$\times$ \\
\bottomrule
\end{tabular*}
\end{table}

In this section we compile a full application, approximate nearest-neighbor search (ANNS) over SIFT1M~\cite{jegou2011SIFT1M}, end-to-end from PyTorch to SIMDRAM on DDR4 with 128 banks. A query is mapped to a 128-bit SimHash~\cite{charikar2002simhash} code by a floating-point random projection and scanned by Hamming similarity against $2^{20}$ codes, whose top $C{=}1000$ are selected and re-ranked by L2 distance, the filter-then-rerank structure that \Cthree{} maps to commodity DRAM by hand. \ToolName{} partitions the torch-mlir input: the projection's $128{\times}128$ iteration space falls far below the offloading threshold and the re-rank iterates over the candidates dynamically (\autoref{sec:overview}), so they stay on the host, an out-of-order x86 core at 2\,GHz with 32\,KB L1 and 1\,MB L2 caches, which also issues the row operations for PUD. CPU-only runs on this core and the same memory system. It uses hardware popcount and scans the codes in their native packed layout, two 64-bit words per code, while the PUD configurations instead keep the codes in vertical layout in DRAM, stored once when the database is built. DRAM energy follows the Micron/DRAMPower model, charging every activation of a row operation as a triple-row activation~\cite{seshadri2017ambit}.

\autoref{tab:ann} enables \ToolName{}'s program-level compilation capabilities one at a time.
\emph{Kernel offload} compiles offloaded operators, here the scan, whose execution time falls from 5.07 to 0.59\,ms. However, two costs that kernel compilation cannot expose still remain. First, each broadcast query is stored and \texttt{XNOR}ed in 128 rows of all banks, although each of its bits is the same for all codes. Second, each step of the selection is a separate small kernel that sends back and transposes operands via host.
\emph{Broadcast specialization} removes the first. As discussed in \autoref{sec:overview}, each query is no longer stored in DRAM rows and each \texttt{XNOR} with a query bit becomes either a \texttt{COPY} or a \texttt{NOT} of the code row, which \Cthree{} derives by hand. The row operations per query fall by 25\%, and broadcast and scan together take 0.2\,ms less.
\emph{Kernel chaining} removes the second by keeping intermediate values shared by chained kernels in DRAM. The data round-trips to the host are eliminated. Small kernels exchange results and operands in-place by sharing 20 rows per bank. The host reads back only their bitmaps, and selection falls from 1.17 to 0.35\,ms.

\section{Related Work}

Despite the growing diversity of PUD architectures, programmability support remains deeply fragmented (\autoref{tab:related}). %
We distinguish \emph{hardware frameworks}, which pair a PUD substrate with an ISA and a library of per-operation $\mu$Programs, from \emph{compilers}, which translate arbitrary programs into primitive sequences. Hardware frameworks such as SIMDRAM~\cite{hajinazar2021simdram} and Proteus~\cite{de2025proteus} expose a fixed set of operations at fixed or runtime-selected bitwidths; any operation outside the library needs a new hand-derived $\mu$Program, and the programmer must express the application in terms of these operations. A PUD compiler instead performs high-level synthesis by deriving the Boolean implementation, row allocation, and schedule of arbitrary operations from the program itself. %
Chopper~\cite{peng2023chopper} and PIMsynth~\cite{guo2025pimsynth} are logic-level compilers for multiple bit-serial PUD substrates. %
Neither lowers tensor programs from ML frameworks, and neither optimizes the Boolean network against the target's row-allocation and copy costs. %

OptiPIM~\cite{liu2025optipim} demonstrates that tensor-level mapping quality improves substantially when the ILP scheduler is informed by lower-level command costs---a key insight that motivates our joint optimization approach---but does not provide a retargetable logic backend. Meta-PIM~\cite{zhang2026meta} compiles C programs for a configurable in-cache fabric, using equality saturation to choose loop tiling and a bit-parallel or bit-serial mode per statement. CINM~\cite{khan2024cinm} is the closest multi-target infrastructure that lowers \texttt{linalg} programs through a device-agnostic MLIR abstraction to UPMEM, CAMs, and crossbar targets, but explicitly leaves each technology-specific backend to the device vendor. \ToolName{} complements rather than competes with CINM, as we provide such a backend for PUD. Because both build on MLIR and enter at \texttt{linalg}, our abstraction is compatible with CINM's. CIM-MLC~\cite{qu2024cim} and OCC~\cite{occ} offer MLIR stacks for analog matrix-vector multiplication in memristive crossbars. %

E-graph-based equality saturation has been applied to FPGA LUT remapping~\cite{hofmann2025eqmap} and ASIC logic synthesis~\cite{chen2024syn}, where cost-driven extraction outperforms greedy rewriting. Their cost models, however, count LUTs, gates, or depth, whereas PUD cost is dominated by destructive operands and row allocation, which our architecture description exposes to extraction. No existing tool jointly optimizes across the tensor and logic levels under a retargetable architecture description.

\section{Conclusion}
\ToolName{} is the first retargetable, end-to-end compilation framework for PUD architectures, treating architectural diversity as a first-class, swappable parameter rather than a fixed compiler assumption. We demonstrate that the same e-graph rewriting and extraction pipeline improves program quality across three fundamentally different PUD substrates (SIMDRAM, PRADA, FCDRAM) by swapping only the cost function, with no architecture-specific heuristics, reducing latency by up to 25\% over decoupled synthesis approaches and by 4--48\% over no rewriting depending on target. At the tensor level, our ILP-based mapping, scheduling, and bitwidth-aware reduction cut memory commands by 11--87\% and produce a valid layout for every shape and bank count, unlike OptiPIM, whose ILP is infeasible for most GEMM configurations and low-bank GEMV, while running 1.06--1.38$\times$ faster.
Feeding the logic level's per-operation costs into the tensor-level mapper makes schedules $1.34$--$3.25\times$ faster than decoupled mapping.
Extending compilation from single kernels to the whole PyTorch program doubles the speedup of vector search on SIFT1M over the CPU baseline, from $2.9\times$ to $5.8\times$.
Together, these results show that treating the architecture description as a shared interface across compilation stages is sufficient to match or exceed hand-optimized, architecture-specific compilers, without sacrificing retargetability.

\begin{acks}
This work was partially funded by the German Research Council (DFG) through the REC\textsuperscript{2} Excellence Cluster (EXC 3035) and the TRR 404 Active 3D (528378584), by the AI competence center ScaDS.AI Dresden/Leipzig (01IS18026A-D) and by the European Research Council (ERC) under the European Union's Horizon Europe research and innovation programme (Consolidator Grant, grant No.~101230491, project COMETH). Views and opinions expressed are however those of the authors only and do not necessarily reflect those of the European Union or the European Research Council Executive Agency. Neither the European Union nor the granting authority can be held responsible for them.

\end{acks}

\bibliographystyle{ACM-Reference-Format}
\bibliography{refs}

\end{document}